# Advancing bioinformatics with large language models: components, applications and perspectives

Jiajia Liu[1,†], Mengyuan Yang[2,†], Yankai Yu[3,†], Haixia Xu[1,†], Tiangang Wang[1], Kang Li[4], Xiaobo Zhou[1,5,6,*]

[1]Center for Computational Systems Medicine, McWilliams School of Biomedical Informatics, The University of Texas Health Science Center at Houston, Houston, Texas, 77030, USA

[2]Department of Cell Biology and Genetics, School of Basic Medical Sciences, Xi'an Jiaotong University Health Science Center, Xi'an, China

[3]School of Computing and Artificial Intelligence, Southwest Jiaotong University, Chengdu, Sichuan 611756, China

[4]West China Biomedical Big Data Center, West China Hospital, Sichuan University, Chengdu, Sichuan 610041, China

[5]McGovern Medical School, The University of Texas Health Science Center at Houston, Houston, TX 77030, USA

[6]School of Dentistry, The University of Texas Health Science Center at Houston, Houston, TX 77030, USA

[†]These authors have contributed equally to this work.

[*] Address correspondence to:

Xiaobo Zhou, Ph.D.

McWilliams School of Biomedical Informatics

The University of Texas Health Science Center at Houston

7000 Fannin St., Houston, TX 77030

Phone: 713-500-3923

Email: Xiaobo.Zhou@uth.tmc.edu

**Abstract**

Large language models (LLMs) are a class of artificial intelligence models based on deep learning, which have great performance in various tasks, especially in natural language processing (NLP). Large language models typically consist of artificial neural networks with numerous parameters, trained on large amounts of unlabeled input using self-supervised or semi-supervised learning. However, their potential for solving bioinformatics problems may even exceed their proficiency in modeling human language. In this review, we will provide a comprehensive overview of the essential components of large language models (LLMs) in bioinformatics, spanning genomics, transcriptomics, proteomics, drug discovery, and single-cell analysis. Key aspects covered include tokenization methods for diverse data types, the architecture of transformer models, the core attention mechanism, and the pre-training processes underlying these models. Additionally, we will introduce currently available foundation models and highlight their downstream applications across various bioinformatics domains. Finally, drawing from our experience, we will offer practical guidance for both LLM users and developers, emphasizing strategies to optimize their use and foster further innovation in the field.

## 1. Introduction

Significant progress has been made in the field of natural language processing with the advent of large language models. Examples of these models include OpenAI's GPT-X [1] and Google's BERT [2] models. These models are transformative because they can understand, generate, and manipulate human language at an unprecedented scale. Vast Large language models are typically trained on datasets that encompass a significant portion of the internet's text, enabling them to learn the complexities of language and context. These models are built upon a neural network architecture called transformers [3]. The transformer architecture revolutionized NLP due to its parallelization, scalability, and ability to capture long-range dependencies in text. Instead of relying on recurrent or convolutional layers, transformers use self-attention mechanisms, as previously described, which allow them to assess the importance of every word in a sentence when understanding context. This innovation is key to their remarkable performance.

The training regimen for large language models comprises two phases: pre-training and fine-tuning. During pre-training, the model is trained on an extensive corpus of text data to acquire proficiency in grammar, factual knowledge, reasoning abilities, and word understanding. Fine-tuning tailors these models for specific tasks like translation, summarization, or question-answering. The adaptability of large language models is a major advantage; they can excel at various NLP tasks without task-specific architectures. However, they have found applications in diverse fields beyond NLP, including biology, healthcare, education, finance, customer service, and more. In particular, there have been many successful applications of large language models in the field of bioinformatics. In this manuscript, we focus on the applications of large language models to several bioinformatic tasks through five areas: DNA level, RNA level, protein level, drug discovery and single-cell analysis. Applications of LLMs in genomics focus on LLMs using DNA sequence; applications of LLMs focus on in transcriptomics using RNA sequence; applications of LLMs in proteomics focus on LLMs using protein sequence; applications of LLMs in drug discovery focus on LLMs using Molecular SMILES (seq) and applications of LLMs in single-cell analysis focus on LLMs using scRNA-seq, scMulti-omics and spatial transcriptomics data (**Figure 1**).

## 2. Understanding the Building Blocks of Large Language Models in Bioinformatics

Building large language models involves several critical components, including tokenization methods, embedding techniques, attention mechanisms, transformer architectures, and the training processes for large-scale models. Each of these elements plays a vital role in enabling the models to process, understand, and generate complex data.

### 2.1 Tokenization and input embedding

Tokenization methods are essential for processing raw input data, breaking it down into smaller, manageable units (tokens) that can be analyzed and processed by models. The choice of tokenization method varies depending on the type of data being handled (**Figure 2a, Table 1**).

In DNA and RNA sequence data, tokenization converts raw nucleotide sequences (A, T, C, G for DNA or A, U, C, G for RNA) into a numerical format suitable for computational models. A common method is one-hot encoding, where each nucleotide is represented as a binary vector with a ‘1’ indicating its position (e.g., [1, 0, 0, 0] for A in DNA), as used in RNA-FM [4] and RNA-MSM [5]. Another widely adopted approach is k-mer tokenization, which segments sequences into overlapping substrings of fixed length ‘k’ (e.g., for k=3, “ATGC” becomes “ATG” and “TGC”). This method is employed in models like DNABERT[6], DNAGPT [7], and RNABERT [8].

Additionally, specialized tokens such as '[IND]' can be introduced to mark the start or end of sequences or to handle unknown characters or gaps, as demonstrated in RNAErnie [9].

In protein language models, the input data primarily includes multiple sequence alignments (MSAs), protein sequences, biomedical/biological text, and cDNA. The basic units of MSAs and protein sequences are amino acids, leading most protein language models to use Single Amino Acid Tokenization, where protein sequences are segmented into individual amino acids. This approach is akin to the k-mers method used for DNA and RNA sequences and is employed in models such as ESM-1b [10], ProtTrans [11], and ProGen [12]. For biomedical and biological text, including general descriptions, conditioning tags in generative models, and resources like Gene Ontology (GO), tokenization methods from natural language processing (NLP) are widely used. Methods like WordPiece Tokenization build vocabulary using frequency-based greedy algorithms and segment text into discrete tokens, as demonstrated in ProtST [13]. For cDNA data, tokenization is similar to that of protein sequences but differs in the basic unit. Instead of amino acids, sequences are tokenized into codons, or triplets of nucleotides, as seen in CaLM [14].

In drug discovery, small molecule drugs account for 98% of commonly used medications [1]. LLMs leverage four main tokenization methods to uncover molecular patterns and drug-target interactions. Atom-level tokenization treats molecules as sequences of individual atoms, analogous to character-level text representation, as seen in K-BERT [15]. MolGPT [16] utilizes a SMILES tokenizer that segments molecular structures into units such as atoms, bond types, and ring markers. A Graph-based VQ-VAE approach enhances this by encoding atoms into context-aware discrete values, distinguishing roles like aldehyde versus ester carbons, based on latent codes derived from a graph-based Vector Quantized Variational Autoencoder (VQ-VAE). This method categorizes atoms into chemically meaningful sub-classes, enriching the molecular vocabulary. Fingerprint tokens, another method, represent molecules through binary or numerical vectors summarizing molecular properties or structural patterns, as seen in SMILES-BERT [17].

Tokenization methods for single-cell profiles include four main strategies. Gene ranking/reindexing-based methods rank genes by expression levels and create tokens using ranked gene symbols or unique integer identifiers, as seen in Geneformer [18] and tGPT [19]. Binning-based methods divide gene expression into predefined intervals, assigning tokens based on the corresponding bin, used in models like scBERT [20] and scGPT [21]. Gene set or pathway-based methods group genes into biologically meaningful sets, such as pathways or Gene Ontology terms, with tokens representing the activation of these sets, exemplified by TOSICA [22]. Patch-based methods segment gene expression vectors into equal-sized sub-vectors, as seen in CIForm [152]. Alternatively, convolutional neural networks (CNNs) can be used to transform the reshaped gene expression matrix into several flattened 2D patches, as demonstrated by scTranSort [23]. Another variation involves reshaping the sub-vectors into a gene expression matrix after segmentation, as employed in scCLIP [24]. In addition to the four methods mentioned above, a more direct approach involves projecting gene expression directly, as seen in models like scFoundation [25], and scMulan [26]. Alternatively, some methods tokenize cells instead of genes, as exemplified by models such as CellPLM [27], ScRAT [28], and mcBERT [29], which utilize cell tokens during model training (**Table 1**). These strategies allow models to capture biological structure and variability, tailoring tokenization to single-cell data characteristics.

After tokenization, embedding converts tokens into continuous vector representations, capturing the semantic relationships between them. Positional encoding represents the token order by adding vectors that encode the relative or absolute positions of tokens in the sequence. The final step involves combining the token embeddings with the positional embeddings to create a unified input embedding, which is then fed into the model for further processing (**Figure 2b**).

### 2.2 Architecture of transformer models

Transformers are the foundational architecture in large language models (LLMs) and consist of two main components: the encoder and the decoder. The encoder takes the input data and processes it in parallel across multiple layers to capture relationships within the sequence. The decoder, on the other hand, generates output sequences based on the encoder's processed information, typically used in tasks like translation or text generation. Each component is built on layers of multi-head attention, add and norm layer, and feed-forward layer (**Figure 2c**).

**Attention Mechanism:** A key innovation of the transformer is the attention mechanism, particularly self-attention [3], which allows the model to weigh the importance of different tokens in a sequence relative to each other. In self-attention, each token computes a score based on how much attention it should pay to other tokens in the sequence. This is done by calculating three key components: Query (Q), Key (K), and Value (V) vectors for each token (**Figure 2d**). The attention score is computed as the dot product between the Query of one token and the Key of another token, followed by a softmax operation to normalize the scores. These scores are then used to weight the

Value vectors, which are aggregated to form the output representation for each token as following [3]:

$$Attention(Q, K, V) = softmax(\frac{QK^T}{\sqrt{d_k}})V \quad (1)$$

**Multi-head attention** extends this idea by running multiple attention mechanisms (or "heads") in parallel. Each attention head processes the input tokens in a slightly different way by using different sets of learned weights for the Q, K, and V vectors. The results of all heads are concatenated and linearly transformed, allowing the model to capture different aspects of relationships between tokens simultaneously. This mechanism enables the model to focus on various parts of the input sequence at once, learning different types of interactions between tokens. For example, in single-cell foundation models, self-attention can help identify important gene interactions by determining which genes (tokens) should focus on each other during processing. In this way, multi-head attention allows the model to capture complex relationships between genes in single-cell RNA-seq data, where multiple aspects of gene expression (such as co-expression patterns or functional relationships) need to be captured simultaneously.

**Add and norm layer:** The add and norm layer performs layer normalization and residual connections, which help stabilize training by ensuring that the output from each layer is added to the input before being normalized. This allows for smoother gradient flow and avoids the vanishing gradient problem.

**Feed-forward layer:** After the attention mechanism, the feed-forward network is a fully connected neural network (**Figure 2e**), helping the model learn complex mappings and capture more abstract representations of the input data.

### 2.3 BERT and GPT models

BERT and GPT stand as two exceptional language models. Both BERT and GPT leverage the transformer architecture, employing attention mechanisms to grasp dependencies within input data.

**BERT (Bidirectional Encoder Representations from Transformers).** BERT is basically an encoder stack of transformer architecture, which was introduced by Google in 2018 [2]. BERT is trained using a bidirectional approach, meaning it considers context from both the left and right of each token during training. This enables BERT to capture richer, more context-aware representations. BERT is typically pre-trained using a masked language model (MLM) task, where random tokens in a sequence are masked, and the model is tasked with predicting them. This bidirectional training allows BERT to better understand the full context of a sequence or biological sentence, such as a cell (**Figure 2f**). For example, scBERT, a single-cell adaptation of BERT, applies this approach to single-cell RNA-seq data. By masking random gene tokens and predicting them during pretraining, scBERT learns complex dependencies and co-expression patterns between genes. This enables it to capture the full transcriptional context of individual cells, improving downstream tasks like cell type classification.

**GPT (Generative Pretrained Transformer).** Introduced by OpenAI [1], GPT is based on a decoder stack of transformer architecture. Unlike BERT, GPT uses a unidirectional training approach, processing the input sequence from left to right (**Figure 2f**). It is trained using autoregressive learning, where each token is predicted based on the previous ones, making it particularly suited for generational tasks. GPT excels in zero-shot learning, where they perform tasks without needing task-specific training data. For example, DNAGPT leverages its pretrained knowledge to perform tasks like predicting DNA motifs or identifying regulatory elements without explicit task-specific training. When prompted with a sequence such as "Find the transcription factor binding motif in the following DNA sequence: AGCTTAGGCC...", DNAGPT can identify or generate plausible motifs based on its understanding of DNA patterns learned during pretraining.

## 3. Foundation models in bioinformatics

### 3.1 Key components of biological foundation models

Foundation models are a category of large-scale, pre-trained models designed to be versatile and adaptable to various downstream tasks. They are built upon several fundamental components that enable their widespread applicability and effectiveness across domains. First, foundation models are trained on extensive and diverse datasets to capture broad, generalizable patterns. In single-cell biology, for example, datasets with millions of cells spanning multiple tissues and conditions are often used. Second, the architecture of foundation models is typically designed for flexibility and scalability. Their architecture, often transformer-based (e.g., GPT and BERT), are specifically designed for flexibility and scalability. Third, Self-supervised learning is a core training strategy for foundation models. By creating tasks such as masked prediction, contrastive learning, or next-token prediction, models can learn representations without requiring labeled data. Fourth, foundation models exhibit multi-task transferability, leveraging a two-step process of pre-training and fine-tuning **(Figure 3)**. During pre-training, these models are trained on large-scale datasets to develop robust generalization capabilities by capturing broad patterns and knowledge. Fine-tuning involves adapting the pre-trained model to specific tasks by exposing it to unique data and additional training. This approach enables foundation models to adjust effectively to diverse applications while maintaining their versatility across a wide range of domains. Last but not least, training foundation models requires significant computational resources, often involving GPU or TPU clusters. Foundation models typically feature billions or even trillions of parameters.

### 3.2 Foundation models in different biological domains

**DNA foundation models.** Currently, DNA sequence-based foundation models are powerful tools that leverage advanced deep learning architectures to analyze and interpret genomic data [30]. These models are built on frameworks like BERT and GPT, which have been adapted for the specific challenges of genomic sequences (**Table 2**). For example, DNABERT [6]is a BERT-based model trained on the human reference genome, enabling it to capture the contextual relationships between nucleotides and perform tasks such as sequence classification and variant prediction. Expanding beyond a single species, Nucleotide Transformer [31] and Genomic Pre-trained Network (GPN) [32] are transformer-based models that incorporate the multiple species reference genomes, providing a broader understanding of genomic diversity. DNABERT-2 [33] takes this a step further by training on multi-species genomic data from 135 species, allowing for cross-species genomic analysis. Similarly, GROVER [34], another BERT-based model, is focused on the human reference genome and is designed for applications such as understanding gene expression and functional genomics. On the other hand, DNAGPT, based on the GPT architecture, is trained not only on the human reference genome but also on reference genomes from nine other species, facilitating tasks such as sequence generation and evolutionary analysis. Together, these DNA sequence-based foundation models represent a leap forward in computational genomics, enabling more accurate predictions, better understanding of genetic variation, and advancements in personalized medicine.

**RNA foundation models.** RNA sequence-based language models, particularly BERT-based and Transformer-based models, have gained significant traction in the analysis of RNA sequences due to their ability to understand the complex patterns and structures of RNA. These models are trained using a wide variety of RNA types, including non-coding RNAs (ncRNAs), coding RNA, and untranslated regions (UTRs), across diverse organisms (**Table 2**). For instance, RNABERT [8], RNA-FM [4], RNA-MSM [5], and UNI-RNA [35] focus on all ncRNA types from a broad range of species, enabling insights into RNA function and interactions. Models like SpliceBERT [36] specialize in coding RNA sequences from 72 vertebrates, while 3UTRBERT [37] is specifically designed for human mRNA transcripts, particularly the 3' untranslated regions. Additionally, UTR-LM [38] focuses on 5' UTR sequences from five species, and RNAErnie [9], a Transformer-based model, covers a wide range of ncRNAs. These models are part of a rapidly growing field aimed at advancing RNA sequence analysis, facilitating the study of RNA biology and its role in various biological processes and diseases. Through the use of these RNA-based language models, researchers can make significant strides in understanding RNA structure, function, and regulatory mechanisms.

**Protein foundation models.** Foundation models for proteins can be directly utilized to obtain high-quality protein embeddings and support various downstream applications. The foundational protein models listed in **Table 2** not only fulfill these requirements but also exhibit unique characteristics. For example, TAPE [39] made a significant contribution by introducing a comprehensive benchmark for protein bioinformatics tasks. ESM-1b [40] applied the transformer architecture of large language models in a highly standardized manner to protein representation learning. This model has since been widely used to generate protein sequence embeddings, and its variants can also be found via the same link provided in **Table 2**. ProtTrans [11], compared to ESM-1b, significantly expanded the model architecture, the number of parameters, and the size of the training dataset. It has been widely adopted as a frozen encoder for protein sequences. ProtGPT2 [41], as its name suggests, extends GPT-2 into the protein domain (with links providing details on the GPT-2 training framework). Recent foundation models like ProtBert [42] and KeAP [43] integrate biomedical text information alongside protein sequences. Notably, KeAP incorporates a knowledge graph to enhance this integration. Both models demonstrate that multimodal fusion within proteomics often produces more expressive features. CaLM [14], on the other hand, represents proteins using cDNA, embedding cross-omics biological information. From the perspective of algorithmic advancements, the integration of multimodal information within a single omics domain, as well as cross-omics data fusion, represents key strategies for constructing unified large-scale biological models.

**Drug discovery foundation models.** It has been postulated that the total number of potential drug like candidates range from $10^{23}$ to $10^{60}$ molecules[44]. Foundation models leverage diverse tokenization strategies, embedding techniques, and pre-training mechanisms to enhance molecular representation learning, facilitating the optimization of various downstream tasks (**Table 2**). For instance, Mol-BERT [45] employs a context-aware tokenizer to encode atoms into chemically meaningful discrete values, although this approach results in an unbalanced atom vocabulary. SMILES-BERT [46], a semi-supervised model incorporating an attention-based Transformer architecture, utilizes datasets such as LogP, PM2, and PCBA-686978 to pre-train the model via a Masked SMILES Recovery (MSR) task. This model demonstrates strong generalization capabilities, enabling its application to diverse molecular property prediction tasks through fine-tuning. Similarly, Mol-GPT [47] facilitates the generation of molecules with specific scaffolds and desired molecular properties by conditioning the generation process on scaffold SMILES strings and property values. Notably, SynerGPT [48] enables a pre-trained GPT model to perform in-context learning of "drug synergy functions", showcasing potential for future advancements in personalized drug discovery. These foundation models developed based on distinct strategies, effectively learn representations from raw sequence data and molecular descriptors. They provide significant insights into the design of small-molecule drugs, drug-drug interactions, and drug-target interactions.

**Single-cell foundation models.** Foundation models in single-cell analysis are revolutionizing the field by offering scalable and versatile solutions for a wide range of tasks, leveraging both cell and gene-level representations (**Table 2**). Models like scBERT [20], tGPT [19], scMulan [26], UCE [49] and CancerFoundation [50] focus on learning robust cell representations, effectively supporting applications such as cell clustering, cell type annotation, batch effect correction, trajectory inference and drug response prediction. These models excel at analyzing heterogeneous cellular populations and uncovering cellular dynamics. In contrast, models like scGPT [21], scFoundation [25], Geneformer [18] GeneCompass [51] and scPRINT [52] combine the ability to learn both cell and gene-level representations. They capture inter-gene relationships and regulatory networks, making them highly effective for tasks such as gene expression profiling, gene regulatory network (GRN) inference, gene perturbation prediction, and drug dose-response prediction. Notably, scGPT can also handle single-cell multi-omics data, facilitating tasks like scRNA-seq and scATAC-seq integration. Another notable model is Nicheformer [53], a foundation model specifically designed for spatial transcriptomics. It focuses on learning cell representations while being highly adaptable to various downstream tasks in spatial transcriptomics, such as spatial label prediction (e.g., cell type, niche, and region labels), niche composition analysis, and neighborhood density prediction. Additionally, Nicheformer can generate joint embeddings of scRNA-seq and spatial transcriptomics data, facilitating the integration of these modalities for a more comprehensive understanding of cellular and spatial interactions.

## 4. Applications of large language models in bioinformatics

Large language models (LLMs) have seen numerous successful applications in bioinformatics, addressing a wide array of tasks across DNA, RNA, protein, drug discovery, and single-cell analysis (**Figure 4**). These applications highlight the adaptability and potential of LLMs in overcoming bioinformatic challenges, enabling deeper insights into complex biological systems and fostering advancements across multiple domains.

### 4.1 Applications of large language models in genomics

The DNA language models take DNA sequence as input, use transformer, BERT, GPT models to solve multiple biological tasks, including genome-wide variant effects prediction, DNA cis-regulatory regions prediction, DNA-protein interaction prediction, DNA methylation (6mA,4mC 5hmC) prediction, splice sites prediction from DNA sequence (**Table 3, Supplementary Figure 1**). A detailed list of DNA language models, their downstream tasks, and the datasets used can be found in **Supplementary Table 1.**

**Genome-wide variant effects prediction.** Genome-wide variant effects prediction is crucial for understanding the role of DNA mutations in species diversity. Genome-wide association studies (GWAS) provide valuable insights but often struggle to identify specific causal variants [30, 54]. The Genome Prediction Network (GPN) [32] addresses this by using unsupervised pre-training on genomic DNA sequences. During this process, GPN predicts nucleotides at masked positions within a 512-bp DNA sequence. This model is particularly effective at predicting rare variant effects, often missed by traditional GWAS methods. Additionally, models like DNABERT, DNABERT-2, and the Nucleotide Transformer also predict variant effects from DNA sequences. These advancements highlight ongoing efforts to better understand how DNA mutations contribute to biological diversity.

**Cis-regulatory regions prediction.** Cis-regulatory sequences, such as enhancers and promoters, play crucial roles in gene expression regulation, influencing development and physiology [55]. However, identifying these sequences remains a major challenge [56]. Pre-trained models like DNABERT, DNABERT-2, GROVER, and DNAGPT have been developed to predict promoter regions and their activities with high accuracy. BERT-Promoter [57] utilizes a pre-trained BERT model for feature representation and SHAP analysis to filter data, improving prediction performance and generalization over traditional methods. Enhancers, which bind transcription factors to regulate gene expression [58, 59], are predicted by iEnhancer-BERT [60], which leverages DNABERT and uses a novel transfer learning approach. This model employs output from all transformer encoder layers and classifies features with a Convolutional Neural Network (CNN). These advancements highlight the growing trend of treating biological sequences as a natural language for computational modeling, offering new tools for identifying cis-regulatory regions and understanding their roles in diseases.

**DNA-protein interaction prediction.** Accurate identification of DNA-protein interactions is crucial for gene expression regulation and understanding evolutionary processes [61]. Several DNA language models, including DNABERT, DNABERT-2, and GROVER, have been developed to predict protein-DNA binding from ChIP-seq data. TFBert [62] is a pre-trained model specifically designed for DNA-protein binding prediction, which treats DNA sequences as natural sentences and k-mer nucleotides as words, allowing effective context extraction. Pre-trained on 690 ChIP-seq datasets, TFBert delivers strong performance with minimal fine-tuning. The MoDNA [63] framework introduces domain knowledge by incorporating common DNA functional motifs. During self-supervised pre-training, MoDNA performs tasks such as k-mer and motif prediction. Pre-training on extensive unlabeled genome data, MoDNA acquires semantic-level genome representations, enhancing predictions for promoter regions and transcription factor binding sites. Essentially, MoDNA functions as a biological language model for DNA-protein binding prediction.

**DNA methylation prediction.** DNA methylation is a key biological process in epigenetic regulation and is linked to various medical conditions and applications, such as metagenomic binning [64]. DNA methylation types depend on the nucleotide where the methyl group attaches [65]. Several models predict DNA methylation with varying accuracy. BERT6mA [66] is designed for predicting 6-methyadenine (6mA) sites, while iDNA-ABT [67], iDNA-ABF [68], and MuLan-Methyl [69] are versatile models predicting various methylation types (6mA, 5hmC, 4mC). iDNA-ABT, a deep learning model, integrates BERT with transductive information maximization (TIM), though it has yet to fully explore feature representation. iDNA-ABF uses a multi-scale architecture, applying multiple tokenizers for diverse embeddings, and MuLan-Methyl employs four transformer-based models (DistilBERT [70], ALBERT[71], XLNet [72], and ELECTRA [73]) to predict methylation sites, enhancing performance through joint model utilization.

**DNA level splice site identification.** Accurate pre-mRNA splicing is essential for proper protein translation, driven by splice site selection. Identifying splice sites is challenging, particularly with prevalent GT-AG sequences [74]. To address this, DNABERT and DNABERT-2 were developed, trained on 10,000 donors, acceptor, and non-splice site sequences from the human reference genome to predict splice sites. DNABERT showed high attention to intronic regions, suggesting the functional role of intronic splicing enhancers and silencers as cis-regulatory elements in splicing regulation. This highlights DNABERT's potential in understanding splicing mechanisms.

### 4.2 Applications of large language models in transcriptomics

The RNA language models take RNA sequences as input, use transformer, BERT, GPT models to solve multiple biological tasks, including RNA 2D/3D structure prediction, RNA structural alignment,, RNA family clustering, RNA splice sites prediction from RNA sequence, RNA N7-methylguanosine modification prediction, RNA 2'-O-methylation modifications prediction, multiple types of RNA modifications prediction, predicting the association between miRNA, lncRNA and disease, identifying lncRNAs, lncRNAs' coding potential prediction, protein expression and mRNA degradation prediction (**Table 3, Supplementary Figure 1**). A detailed list of RNA language models, their downstream tasks, and the datasets used can be found in **Supplementary Table 1.**

**Secondary structure prediction.** RNA secondary structure prediction is a major challenge for RNA structural biologists, with models holding potential for RNA-targeting drug development [75]. Several RNA language models, such as RNABERT [8], RNA-MSM [5], RNA-FM [4], and UNI-RNA [35], have been developed to predict RNA structures with varying sophistication. RNABERT uses BERT architecture to predict structural features like base-pairing and stem loops. RNA-MSM integrates sequence and structural information to predict local and long-range folding patterns. RNA-FM focuses on RNA folding, stability, and energetics, including pseudoknots. UNI-RNA combines sequence and structure predictions across various RNA types. These models advance RNA structure prediction by applying deep learning and advanced techniques to improve understanding of RNA folding and function.

**RNA splicing prediction**. RNA splicing is crucial for gene expression in eukaryotes, and advancements have been made in sequence-based splicing modeling through models like

SpliceBERT [36] and UNI-RNA [35]. SpliceBERT, based on BERT, is trained to predict RNA splicing events by capturing long-range dependencies, identifying splice sites, and predicting alternative splicing events. UNI-RNA, a more generalized model, integrates multiple RNA tasks, including splicing, and combines sequence and structural data to predict splicing regulatory elements and interactions with splicing factors. These models enhance the understanding of RNA splicing, gene regulation, and its role in diseases, providing powerful tools for studying splicing defects and mutations.

**lncRNAs identification and lncRNAs' coding potential prediction.** Long non-coding RNAs (lncRNAs) play significant regulatory roles in cancer and diseases, and their small Open Reading Frames (sORFs), once thought weak in protein translation, are now known to encode peptides [76]. Identifying lncRNAs with sORFs is crucial for discovering new regulatory factors. LncCat [77] addresses this challenge by using category boosting and ORF-attention features, including BERT for peptide sequence representation, to improve prediction accuracy for both long ORF and sORF datasets. It demonstrates effectiveness across multiple species and Ribo-seq datasets in identifying lncRNAs with sORFs. In predicting translatable sORFs in lncRNAs (lncRNA-sORFs), LSCPP-BERT [78] is a novel method designed for plants, leveraging pre-trained transformer models for reliable coding potential prediction. LSCPP-BERT is poised to impact drug development and agriculture by enhancing understanding of lncRNA coding potential.

**RNA–RBP interactions prediction.** RNA sequences differ from DNA sequences by a single base (thymine to uracil), maintaining largely congruent syntax and semantics. BERT's versatility extends to Cross-linking and Immunoprecipitation data, particularly in predicting RNA-binding protein (RBP) binding preferences. BERT-RBP [79] is a model pre-trained on a human reference genome, designed to forecast RNA-RBP interactions. It outperforms existing models when tested on eCLIP-seq data from 154 RBPs and can identify transcript regions and RNA secondary structures based on sequence alone. BERT-RBP demonstrates BERT's adaptability in biological contexts and its potential to advance RNA-protein interaction understanding.

**RNA-RNA interaction prediction**. RNA–RNA interactions occur between various RNA species, including long non-coding RNAs, mRNAs, and small RNAs (e.g., miRNAs and lncRNAs), driven by complementary sequences, secondary structures, and other motifs [80]. Accurate prediction of these interactions provides insights into RNA-mediated regulation, enhancing understanding of biological processes like gene expression, splicing, and translation. RNAErnie, used for this purpose, employs a TBTH architecture combining RNAErnie with a hybrid network (CNN, Bi-LSTM, and MLP) to predict RNA–RNA interactions. This approach demonstrates RNAErnie's potential in advancing RNA-based regulatory network studies.

**RNA modification prediction.** Post-transcriptional RNA modifications, such as N7-methylguanosine (m7G) and 2'-O-methylation (Nm), regulate gene expression and are linked to diseases [76, 81]. Identifying modification sites is essential but challenging due to the high cost and time required by experimental methods. Computational tools like BERT-m7G [82] and Bert2Ome [83] address this issue. BERT-m7G uses a stacking ensemble approach to identify m7G sites directly from RNA sequences, offering an efficient, cost-effective alternative. Bert2Om combines BERT and CNN to predict 2'-O-methylation sites, outperforming existing methods across datasets and species. These tools enhance the accuracy, scalability, and efficiency of RNA modification site identification, advancing research into RNA modifications and their roles in gene regulation and disease.

**Protein expression and mRNA degradation prediction.** mRNA vaccines are a cost-effective, rapid, and safe alternative to traditional vaccines, showing high potency [84]. These vaccines work by introducing mRNA that encodes a viral protein. CodonBERT [85] is a model specifically designed for mRNA sequences to predict protein expression. It uses a multi-head attention transformer architecture and was pre-trained on 10 million mRNA sequences from various organisms. This pre-training enables CodonBERT to excel in tasks like protein expression and mRNA degradation prediction. Its ability to integrate new biological information makes it a valuable tool for mRNA vaccine development. CodonBERT surpasses existing methods, optimizing mRNA vaccine design and improving efficacy and applicability in immunization. Its strength in predicting protein expression enhances mRNA vaccine development efficiency and effectiveness.

**5' UTR-based mean ribosome loading prediction and mRNA subcellular localization prediction**. The 5' UTR sequence plays a critical role in regulating translation efficiency. RNA sequence models like 3UTRBERT, UNI-RNA, UTR-LM, RNA-FM, and Nucleotide Transformer have been developed to predict key features of the 5' UTR, focusing on ribosome loading efficiency and mRNA localization. These models use Transformer-based architecture to analyze sequence patterns, motifs, and structural elements. For example, 3UTRBERT [37] and RNA-FM [4] predict ribosome loading efficiency, identifying regions likely to recruit ribosomes for translation initiation. UTR-LM [38], UNI-RNA [35], and Nucleotide Transformer [31] predict mRNA subcellular localization, determining where mRNA will localize in the cell (cytoplasm, ribosomes, or nucleus), which is crucial for regulating mRNA stability and translation. Together, these models provide valuable insights into gene expression, translation control, and RNA localization, advancing molecular biology research.

### 4.3 Applications of large language models in proteomics

Protein is an indispensable molecule in life, assuming a pivotal role in the construction and sustenance of vital processes. As the field of protein research advances, there has been a substantial surge in the accumulation of protein data [86]. In this context, the utilization of large language models emerges as a viable approach to extract pertinent and valuable information from these vast reservoirs of data. Several pre-trained protein language models (PPLMs) have been proposed to learn the characteristic representations of proteins data (e.g., protein sequences, gene ontology annotations, property descriptions), then applied to different tasks by fine-tuning, adding or altering downstream networks, such as protein structure, post-translational modifications (PTMs), and biophysical properties, which align with corresponding downstream tasks like secondary structure prediction, major PTMs prediction, and stability prediction [87, 88].

Even though antibodies are classified as proteins, the datasets of antibodies and subsequent tasks differ significantly from those of proteins. Through the establishment and continuous updates of the Observed Antibody Space (OAS) database [89], a substantial amount of antibody sequence data has become available, which can be utilized to facilitate the development of pre-trained antibody large language models (PALMs). PALMs primarily delve into downstream topics encompassing therapeutic antibody binding mechanisms, immune evolution, and antibody discovery, which correspond to tasks like paratope prediction, B cell maturation analysis, and antibody sequence classification (**Table 3, Supplementary Figure 2**).

In this section, some of the popular protein-related large language models of recent years are introduced, as well as corresponding important downstream tasks. It is important to emphasize that both PPLM and PALM are capable of handling not only the downstream tasks introduced in this section. For further details, additional information can be referenced within **Supplementary Table 2**.

**Secondary structure and contact prediction.** Protein structure is critical to its function and interactions [90]. However, traditional experimental techniques for protein structure analysis are time-consuming and labor-intensive. With the rise of deep learning, large language models have demonstrated significant advantages in computational efficiency and prediction accuracy for protein structure prediction [91]. MSA Transformer [92] introduces a protein language model that processes MSAs using a unique mechanism of interleaved row and column attention. Trained with a MLM objective across diverse protein families, it outperformed earlier unsupervised approaches and showed greater parameter efficiency than previous models. Drawing on insights from BERT, large parameter models tend to achieve better performance for predicting secondary structures and contacts. Few models have more parameters than the largest models in ProtTrans [11], which includes a series of autoregressive models (Transformer-XL [93], XLNet [72]) and four encoder (BERT [2], Albert [71], Electra [73], T5 [94]) trained on datasets like UniRef [95] and BFD [96], comprising up to 393 billion amino acids. Model sizes vary from millions to billions of parameters. Notably, ProtTrans made a significant breakthrough in per-residue predictions.

**Protein sequence generation.** Protein sequence generation holds significant potential in drug design and protein engineering [97]. Using machine learning or deep learning, generated sequences aim for good foldability, stable 3D structures, and specific functional properties, such as enzyme activity and antibody binding. The development of large language models, combined with conditional models, has greatly advanced protein generation [98]. ProGen [12] incorporates UniprotKB keywords as conditional tags, covering over 1,100 categories like 'biological process' and 'molecular function'. Proteins generated by ProGen, assessed for sequence similarity, secondary structure, and conformational energy, exhibit desirable structural properties. In 2022, ProtGPT2 [41] inspired by GPT-x was developed. ProtGPT2-generated proteins show amino acid propensities like natural proteins. Prediction of disorder and secondary structure reveals that 88% of these proteins are globular, resembling natural sequences. Employing AlphaFold [99, 100] on ProtGPT2 sequences produces well-folded, non-idealized structures with unique topologies not seen in current databases, suggesting ProtGPT2 has effectively learned “protein language”.

**Protein function prediction.** Proteins are essential in cellular metabolism, signal transduction, and structural support, making their function critical for drug development and disease analysis. However, predicting and annotating protein functions is challenging due to their complexity. PPLMs offer effective solutions to these challenges [101, 102]. ProtST [103] introduced a multimodal framework combining a PPLM for sequences and a biomedical language model (BLM) for protein property descriptions. Through three pre-training tasks, unimodal mask prediction, multimodal representation alignment, and multimodal mask prediction, the model excels in tasks like protein function annotation, zero-shot classification, and functional protein retrieval from large databases. While most methods focus on increasing model parameters to improve performance, CaLM [14] introduces an alternative representation, the cDNA sequence, akin to an amino acid sequence, as input. The core idea lies in the relationship between synonymous codon usage and protein structure [104], and the information encoded in codons is no less than that of amino acids. Experimental results demonstrate that even with a small parameter language model, using cDNA sequences as input enhances performance in tasks such as protein function prediction, species recognition, prediction of protein and transcript abundance, and melting point estimation.

**Major post-translational modification prediction.** Post-translational modifications (PTMs) are chemical changes, such as phosphorylation, methylation, and acetylation, that alter protein structure and function after translation. PTMs influence protein stability, localization, interactions, and function, making their study crucial for disease diagnosis and therapeutic strategies [105, 106]. Language models can effectively predict PTMs and related tasks like signal peptide prediction.

ProteinBERT [42] , with only ~16M parameters, is not large enough but performs well due to its inclusion of Gene Ontology (GO) annotation tasks. By incorporating GO interactions with protein sequences, ProteinBERT achieves strong performance on PTM prediction and other protein property benchmarks, outperforming models with larger parameter sizes.

**Evolution and mutation prediction.** Protein evolution and mutation drive functional diversity, aiding adaptation to environmental changes and offering insights into protein function origin, which can inform drug development and disease treatment [107, 108]. UniRep [109], built on the LSTM architecture, was trained on the UniRef50 [95] and excelled in tasks like remote homology detection and mutation effect prediction. ESM-1b [40] , a deep transformer model trained on 250 million sequences, with 33 layers and 650 million parameters, captures essential protein sequence patterns through self-supervised learning. ESM-1b is also integral to frameworks like PLMSearch [110] and DHR [111] , which enable fast, sensitive homology searches. PLMSearch uses supervised training, while DHR relies on unsupervised contrastive learning and enhances structure prediction models like AlphaFold2 [100].

**Biophysical properties prediction.** Biophysical properties of proteins, such as fluorescence and stability landscapes [112], are crucial for understanding protein folding, stability, and conformational changes, with significant implications for drug design, protein engineering, and enzyme engineering. Deep learning advancements have enabled more accurate prediction of these properties using PPLMs. TAPE benchmark [39] established standardized tasks for evaluating protein, including fluorescence and stability landscape prediction. In 2022, PromptProtein [113], a prompt-based pre-trained model, incorporated multi-task pre-training and a fine-tuning module to improve task-specific performance. It outperformed existing methods in function and biophysical properties prediction, demonstrating substantial gains in predictive accuracy.

**Protein-protein interaction and binding affinity prediction.** Protein-protein interactions (PPIs) are crucial for biological functions, and their prediction is also vital for drug discovery and design. PPLMs provide efficient, accurate predictions of PPI types and binding affinities [114, 115]. KeAP model [43], like ProtST, aims to integrate fine-grained information beyond OntoProtein [116]. KeAP uses a triplet format (Protein, Relation, Attribute) as input, processed by encoders and a cascaded decoder based on the Transformer architecture. Using MLM for pre-training, KeAP employs a cross-attention fusion mechanism to capture detailed protein information, achieving superior performance on tasks such as PPI identification and binding affinity estimation.

**Antigen-receptor binding and antigen-antibody binding prediction.** Antigen proteins are processed into neoantigen peptides that bind to the Major Histocompatibility Complex (MHC), forming pMHC complexes. These complexes are presented to T-cells, stimulating antibody production by B-cells, which triggers an immune response [117] . Predicting peptide binding to MHC molecules is a key focus of language models in this process [118, 119]. MHCRoBERTa [120] uses a pretrained BERT model to predict pMHC-I binding by learning the biological meaning of amino acid sequences. BERTMHC [121], trained on 2,413 MHC–peptide pairs, focuses on pMHC-II binding prediction, filling a gap in this area.

Another goal is predicting the binding specificity of adaptive immune receptors (AIRs), particularly TCRs. TCR-BERT [122] learns TCR CDR3 sequences to predict antigen specificity but lacks the ability to model the interaction between TCR chains. SC-AIR-BERT [123] addresses this by pre-training a model that outperforms others in TCR and BCR binding specificity. Additionally, the Antiformer [124] integrates RNA-seq and BCT-seq data in a graph-based framework to improve antibody development. In antibody modeling, three recent models focus on unique tasks. AbLang [125], built on RoBERTa [126], excels at restoring lost residues during sequencing and outperforms other models in accuracy and efficiency. AntiBERTa [127] understands antibody "language" through tasks like predicting immunogenicity and binding sites.

EATLM [128] , with its unique pre-training tasks (Ancestor Germline Prediction and Mutation Position Prediction), contributes a reliable benchmark for antibody language models.

### 4.4 Applications of large language models in drug discovery

Drug discovery is an expensive and long-term process that exhibits a low success rate. During the early stages of drug discovery, computer-aided drug discovery, employing empirical or expert knowledge algorithms, machine learning algorithms, and deep learning algorithms, serve to accelerate the generation and screening of drug molecules and their lead compounds [129-131]. It speeds up the entire drug discovery process, especially the development of small molecule drugs. Among commonly used medications, small molecule drugs can account for up to 98% of the total [132]. The structure of small molecule drugs exhibits excellent spatial dispersibility, and their chemical properties determine their good drug-like properties and pharmacokinetic properties [133]. With the development of deep learning and the proposal of large language models, it has become easy to apply these methods to discover hidden patterns of molecules and interactions between molecules for drugs (such as small molecules) and targets (such as proteins and RNA) that can be easily represented as sequence data. The Simplified Molecular-Input Line-Entry System (SMILES) string and chemical fingerprint are commonly used to represent molecules. Additionally, through the pooling process of graph neural networks(GNN), small molecules can be transformed into sequential representations [134]. With the protein sequence, large language models can engage in drug discovery through various inputs. Within this section, key tasks within the early drug discovery process that have effectively leveraged large language models will be introduced (**Table 3, Supplementary Figure 3**). A detailed list of drug discovery language models, their downstream tasks, and the datasets used can be found in **Supplementary Table 3**.

**Drug-like molecular properties prediction**. In drug discovery, significant focus is placed on properties like ADMET and PK to develop more effective, accessible, and safe drugs[135, 136]. Large language models (LLMs) are used for molecular property prediction, including these properties. Since molecular SMILES representations are consistent, models can be easily improved and fine-tuned for specific tasks based on researchers' requirements. SMILES-BERT [17] departed from the usage of knowledge-based molecular fingerprints as input. Instead, it adopted a representation method where molecules were encoded as SMILES sequences and employed as input for both pre-training and fine-tuning within a BERT-based model. This novel approach yielded superior outcomes across various downstream molecular property prediction tasks, surpassing the performance of previous models reliant on molecular fingerprints. ChemBERTa [137] is a BERT-based model that focuses on the scalability of large language models, exploring the impact of pre-training dataset size, tokenizer, and string representation. Subsequently, ChemBERTa-2[138] improved upon ChemBERTa by using a larger dataset of 77 million compounds from PubChem, enhancing its ability to learn from diverse chemical structures. It also integrates advanced self-supervised learning techniques and fine-tuning strategies, resulting in better generalization performance across various downstream tasks. K-BERT [15] stands out by using three pre-training tasks: atom feature prediction, molecular feature prediction, and contrastive learning. This approach enables the model to understand the essence of SMILES representations, resulting in exceptional performance across 15 drug datasets, highlighting its effectiveness in drug discovery. Given the importance of graph neural networks in the development of molecular pre-training models, Mole-BERT [139] introduces atom-level Masked Atoms Modeling (MAM) task and graph-level Triplet Masked Contrastive Learning (TMCL) task. These tasks enable the network to acquire a comprehensive understanding of the "language" embedded within molecular graphs. By adopting this approach, the network demonstrates exceptional performance across eight downstream tasks, showcasing its adaptability and effectiveness in diverse applications.

**Drug-like molecules generation.** It is very difficult to chase the full coverage of the enormous drug-like chemical space (estimated at more than $10^{63}$ compounds), and traditional virtual screening libraries usually contain less than $10^{7}$ compounds and are sometimes not available. In such circumstances, the utilization of deep learning methods to generate molecules exhibiting drug-like properties emerges as a viable approach [140, 141]. Inspired by the generative pre-training model GPT, MolGPT [16] model was introduced. In addition to performing the next token prediction task, MolGPT incorporates an extra training task for conditional prediction, facilitating the capability of conditional generation. Beyond its capacity to generate innovative and efficacious molecules, the model has demonstrated an enhanced ability to capture the statistical characteristics within the dataset.

**Drug-target interaction predictions.** The investigation of Drug-Target Interaction (DTI) holds paramount significance in the realm of drug development and the optimization of drug therapy. Understanding drug-target interactions aids in pharmaceutical design, accelerates drug development, and reduces time and resource costs in lab experimentation and trial-and-error methods [142, 143]. During the exploration of DTI, diligent focus is placed on the prediction of drug-target binding affinity. DTI-BERT employs a fine-tuned ProtBERT [144] model to process protein sequences and applies discrete wavelet transform to drug molecular fingerprints.. TransDTI [145] is a multi-class classification and regression workflow. This model not only uses fine-tuned SMILES-BERT to extract drug features, but also expands the selection of fine-tuned large protein models. After acquiring potential representations of drug-target pairs, the authors subject the representations to downstream neural networks for the completion of a multi-classification task. Additionally, The Chemical-Chemical Protein-Protein Transferred DTA (C2P2) [146] method uses pre-trained protein and molecular large language models to capture the interaction information within molecules. Given the relatively limited scale of the DTI dataset, C2P2 leverages protein-protein interaction (PPI) and chemical-chemical interaction (CCI) tasks to acquire knowledge of intermolecular interactions and subsequently transfer this knowledge to affinity prediction tasks [147]. It is worth highlighting that in scenarios involving the docking or when emphasizing the spatial structure of a complex, methodologies incorporating 3D convolution networks, point clouds-based networks, and graph networks are often employed [148-151]. In situations where the molecular structure is unknown, but the sequence is available, the prediction of DTI using large-scale models still holds significant promise.

**Drug synergistic effects predictions.** Combination therapy is common for complex diseases like cancer, infections, and neurological disorders, often surpassing single-drug treatments. Predicting drug pair synergy, where combining drugs boosts therapeutic effects, is vital in drug development. However, it’s challenging due to many drug combinations and complex biology [152, 153]. Various computational methods, including machine learning, help predict drug pair synergy. Carl Edwards et al. introduced SynerGPT [48], which is based on GPT trained to in-context learn drug synergy functions without relying on domain-specific knowledge. Wei Zhang et al. [154] introduced DCE-DForest [154], a model for predicting drug combination synergies. It uses a pretrained drug BERT model to encode the drug SMILES and then predicts synergistic effects based on the embedding vectors of drugs and cell lines using the deep forest method. Mengdie Xua et al. [155] utilized a fine-tuned pre-trained large language model and a dual feature fusion mechanism to predict synergistic drug combinations. Its input includes hashed atom pair molecular fingerprints of drugs, SMILES string encodings, and cell line gene expressions. They conducted ablation analyses on the dual feature fusion network for drug-drug synergy prediction, highlighting the significant role of fingerprint inputs in ensuring high-quality drug synergy predictions.

**4.5 Applications of large language models in single-cell analysis**

Large language models have demonstrated significant applications in single-cell analysis, including cell-level tasks such as identifying cell types, determining cell states, and discovering novel cell populations; gene-level tasks like inferring gene regulatory networks; and multi-omics tasks, such as integrating single-cell multi-omics (scMulti-omics) data (**Supplementary Figure 4**). Additionally, this section will explore emerging language models based on spatial transcriptomics (**Table 3).** A detailed list of single-cell large language models, their downstream tasks, and the datasets used can be found in **Supplementary Table 4.**

**Cell-level tasks.** Cell-level tasks, such as cell clustering, cell type annotation, novel cell type discovery, batch effect removal and trajectory inference, are critical in single-cell analysis. These tasks often rely on cell representations learned during pretraining, which are subsequently fine-tuned for different tasks. Single-cell language models derive cell representations in two primary ways. The first method utilizes a special class token (<cls>) appended to the input sequence; its embedding is updated through the transformer layers, and the final embedding at the <cls> position serves as the cell representation. The second method generates a cell embedding matrix from the model output, where each row represents a specific cell. Both approaches facilitate downstream tasks, as demonstrated by TOSICA [22], which uses the <cls> token to predict cell type probabilities using the whole conjunction neural network cell type classifier to annotate single cells, and iSEEK [156], which generates cell embedding for cell clustering, cell type annotation, and developmental trajectory exploration. Models like scBERT [20] and UCE [49] leverage multi-head attention mechanisms to extract information from diverse representation subspaces, discerning subtle differences between novel and known cell types. Their large receptive fields capture long-range gene-gene interactions, enabling comprehensive characterization of novel cellular states. Addressing batch effects, which arise from variations in species, tissues, operators, and experimental protocols, remains a significant challenge in single-cell analysis. Large language models, pretrained on extensive datasets, utilize attention mechanisms to incorporate prior biological knowledge, enabling batch-insensitive data annotation. Without relying on explicit batch information, models such as CIForm [152] have demonstrated effectiveness in both intra-dataset and inter-dataset scenarios. They handle annotations across diverse species, organs, tissues, and technologies while also supporting the integration of reference and query data from various sequencing platforms or studies. This capability allows them to address batch effects in single-cell analysis. Drug response or sensitivity prediction is a classification task akin to cell type annotation, where a classifier is appended to the learned cell embeddings to predict whether a cell will respond to or exhibit sensitivity to a specific drug. Models like scFoundation [25] and CellLM [157] effectively utilize this approach, leveraging the robust cell representations learned during pretraining to enhance prediction accuracy.

**Gene-level tasks.** Gene-level tasks, such as gene expression prediction, gene regulatory network (GRN) inference, gene perturbation prediction, and drug dose-response prediction, are integral to understanding single-cell transcriptomics. Self-attention mechanisms have transformed deep learning by enabling context-aware models that prioritize relevant elements in large input spaces. These models, particularly transformers, are well-suited for modeling the context-dependent dynamics of gene regulatory networks. By focusing on key interactions, transformers can effectively capture the complexities of regulatory relationships, such as the attention matrix in Geneformer [18] and scGPT [21] reflect which genes that gene pays attention to and which genes pay attention to that gene, aiding to infer gene regulation network. Geneformer is pretrained on a vast repository of single-cell transcriptomes to learn gene relationships for diverse downstream applications, including predicting dosage-sensitive disease genes, identifying downstream targets, forecasting chromatin dynamics, and modeling network dynamics. In addition, after pretraining and fine-tuning, single-cell language models output gene embeddings that can be utilized for functional analysis of scRNA-seq data. For instance, scGPT [21] serves as a generalizable feature extractor leveraging zero-shot learning, enabling applications in gene expression prediction and genetic perturbation prediction. Similarly, in scFoundation [25], zero-expressed genes and masked genes are combined with the output from the transformer-based encoder. This combined information is then input into the decoder and projected to gene expression values through a multilayer perceptron (MLP). The gene context expression is employed to formulate a cell-specific gene graph, facilitating the prediction of perturbations using the GEARS [158] model. It is worth noting that genes have a lot of prior knowledge that can be used to enhance many gene-level tasks. For example, GeneCompass [51] incorporates four types of biological prior knowledge, including GRN, promoter information, gene family annotation and gene-co-expressed relationship, making it capable for various gene tasks.

**scMulti-omics tasks.** Studying single-cell multi-omics data requires integrating diverse information from genomics, transcriptomics, epigenomics, and proteomics at the single-cell level. The adaptability, generalization capabilities, and feature extraction strengths of large language models make them effective in addressing challenges such as feature variance, data sparsity, and cell heterogeneity inherent in single-cell multi-omics datasets. scMulti-omics integration can be viewed as a specialized form of batch effect removal. For example, scGPT [21] treats each modality as a distinct batch and incorporates a special modality token to represent input features (such as genes, regions, or proteins) associated with each modality. This approach helps the transformer model balance attention across modalities, preventing overemphasis on intra-modality features while integrating inter-modality relationships effectively. Another approach involves processing different modalities through separate transformers before projecting their embeddings into a common latent space. Models like scMVP [159] use mask attention-based encoders for scRNA-seq data and transformer-based multi-head self-attention encoders for scATAC-seq. By aligning variations between different omics in this latent space, scMVP captures joint profiling of scRNA-seq and scATAC-seq, achieving paired integration where gene expression and chromatin accessibility are studied within the same cells. Graphs are increasingly recognized as powerful tools for characterizing feature heterogeneity in scMulti-omics integration. For example, DeepMAPS [160] leverages graph transformers to construct cell and gene graphs, learning both local and global features that build cell-cell and gene-gene relationships for data integration, inference of biological networks from scMulti-omics data and cell-cell communication.

Recent advances in sequencing technologies that capture multiple modalities within the same cell have enabled the development of computational tools for cross-modality prediction. One approach involves training large language models on paired datasets to predict one modality from another. For instance, scTranslator [161], pre-trained on paired bulk and single-cell data, fine-tunes to infer protein abundance from scRNA-seq data by minimizing the mean squared error (MSE) between predicted and actual protein levels. Another strategy leverages graph learning with prior knowledge to model feature relationships. For example, scMoFormer [162] can not only translate gene expression to protein abundance, but is also applicable to multi-omics predictions, including protein abundance to gene expression, chromatin accessibility to gene expression, gene expression to chromatin accessibility using graph transformers. Taking protein prediction task as an example, scMoFormer constructs cell-gene graph, gene-gene graph, protein-protein graph, and gene-protein graph based on gene expression profiles and prior knowledge from STRING database [163]. Each modality has a separate transformer to learn the global information that may not be included in prior knowledge. Message-passing graph neural networks (GNNs) link nodes across various graphs, while transformers are employed to precisely map gene expression to protein abundance.

**Spatial transcriptomics tasks.** The rapid development of single-cell and spatial transcriptomics has advanced our understanding of cellular heterogeneity and tissue architecture. Spatial transcriptomics retains cells' native spatial context, enabling insights into cellular interactions. Large language models address the challenge of high-dimensional spatial data analysis by integrating spatial and molecular information, enhancing tissue-specific pattern interpretation. For example, Nicheformer [53] is the latest large language model in spatial transcriptomics. It integrates extensive spatial transcriptomics and single-cell transcriptomics data, leveraging metadata across multiple modalities, species, and sequencing technologies. By doing so, Nicheformer is capable of learning joint information from single-cell and spatial transcriptomics, enabling the resolution of various spatial prediction tasks even with limited data. Spaformer [164] is another transformer-based model designed for spatial transcriptomics data. Spaformer is designed to address two key challenges: how to encode spatial information of cells into a transformer model and how to train a transformer to overcome the sparsity of spatial transcriptomics data, enabling data imputation. Spatial transcriptomics, as one of the most popular technologies in recent years, focuses on integrating single-cell resolution gene expression data with tissue spatial information to reveal spatial relationships and functional characteristics among cells. However, large language models (LLMs) specifically designed for spatial transcriptomics are still in their early stages of development. The creation of these models faces unique challenges, such as effectively integrating high-dimensional gene expression data with complex spatial information and addressing the sparsity and irregularity of the data.

In addition to the single-cell large language models discussed above, another category of single-cell prediction models leverages natural language, utilizing textual data such as human-readable descriptions of gene functions and biological features to support various single-cell analyses. For example, GPT-4 [165] leverages its strong contextual understanding for interpreting high-dimensional single-cell analysis for accurate cell type annotation. GenePT [166] utilizes OpenAI's ChatGPT text embedding to classify gene properties and cell types effectively. More and more models demonstrate that natural language pretraining can significantly boost performance on single-cell downstream tasks, including cell generation [167], cell identity (e.g., cell type, pathway, and disease information) [167-171], and gene enrichment analysis [169]. These models demonstrate significant potential in advancing single-cell analysis by integrating natural language processing techniques. However, the reliance on textual data may constrain performance in less-annotated or novel datasets.

## 5. Conclusion and Suggestions on large language models in bioinformatics

### 5.1 Summary of large language models in bioinformatics

Large language models (LLMs) have catalyzed transformative progress across biological disciplines, including genomics, transcriptomics, proteomics, drug discovery, and single-cell analysis. These models, trained on vast datasets, address challenges like the sparsity, high dimensionality, and heterogeneity of biological data while capturing the complexity of sequence relationships. Tokenization methods are pivotal, converting sequences into manageable formats, such as, for genomics and transcriptomics, k-mer encoding is prevalent, segmenting DNA/RNA sequences into overlapping units. In proteomics, amino acid residue-based tokenization captures protein structure and function. These preprocessing strategies enable LLMs to interpret biological language effectively.

Representation learning allows LLMs to uncover contextual and hierarchical relationships within biological data, forming the basis for various downstream applications. These tasks can be grouped into four primary categories: 1) Classification/Prediction Tasks: Examples include identifying functional genomic elements (e.g., promoters, enhancers), predicting protein structures and interactions, and cell type annotation in single-cell data. 2) Generation Tasks: LLMs can create biologically relevant sequences, such as gene expression imputation and synthetic DNA, RNA, or protein sequences, aiding in vaccine development or enzyme engineering. 3) Interaction Tasks: These involve modeling interactions like drug-target binding, cell-cell interaction, protein-protein interactions, or cross-omics relationships (e.g., gene expression to protein abundance). 4) Transfer Learning Tasks: Pretrained LLMs, such as DNABERT and scGPT, are fine-tuned for specific applications, including single-cell data annotation or predicting RNA modifications like N6-methyladenosine sites. Despite their capabilities, challenges persist. Biological data often exhibit sparsity, as seen in single-cell and spatial transcriptomics, and irregularity due to sequencing errors or noise. To address this, LLMs must effectively integrate multi-modal data, balance computational efficiency, and ensure interpretability of their outputs. As foundational models evolve, their ability to unify diverse biological datasets into a single framework for prediction, generation, interaction, and transfer learning tasks will continue to reshape our understanding and applications of biological systems.

### 5.2 Guidance on how to use and develop LLMs in practice

Large Language Models offer immense potential in bioinformatics and other fields, but their effective utilization and development require distinct approaches for end-users and developers (**Figure 5**).

For LLM users, the process begins by clearly defining the research domain and task, specifying the relevant omics level (e.g., genomics, transcriptomics, proteomics) and identifying whether the objective involves classification or prediction, generation, interaction, or transfer learning. A well-defined objective streamlines the selection of appropriate models and workflows. Next, users should choose models pretrained on data relevant to their domain, as detailed in **Table 2**, which includes information on foundation models, their training data types, and availability. For instance, DNABERT is ideal for genomics tasks, while scGPT is tailored for single-cell analysis. Additionally, users must assess computational requirements and ensure compatibility with their dataset size and complexity. Proper data preparation is critical, including aligning data with model requirements, addressing missing values, and incorporating metadata like cell types or genomic regions. **Table 1** provides common tokenization methods for reference. To leverage transfer learning, users can fine-tune foundation models listed in **Table 2** for their specific dataset, optimizing performance through hyperparameter tuning, early stopping, and cross-validation. Alternatively, users can utilize predeveloped models listed in **Supplementary Tables 1–4** for similar tasks to obtain results directly. Finally, rigorous evaluation using metrics like accuracy, precision, and recall is essential, complemented by interpretation tools such as attention maps or feature embeddings to extract meaningful biological insights (**Figure 5**).

For LLM developers, it is essential to **first** understand domain-specific challenges to address issues like sparsity, heterogeneity, and high dimensionality. For example, single-cell and spatial transcriptomics datasets often suffer from sparsity and noise, necessitating innovative solutions in model architecture. **Second,** developers should choose or develop tokenization strategies tailored to biological data. For instance, k-mer encoding works well for DNA/RNA sequences, while gene ranking-based tokenization is effective for scRNA-seq data. Exploring hybrid tokenization can enhance cross-modal understanding. **Third,** in model development, developers should employ or design novel transformer structures. For example, scBERT utilizes Performer to improve scalability. Incorporating knowledge-based information into model training can further enhance performance. For instance, GeneCompass integrates four types of biological prior knowledge including GRNs, promoter information, gene family annotation, and gene co-expression relationships, making it versatile for various gene-related tasks. Similarly, basic protein language models, which are often limited to MSA and protein sequences, can be improved by incorporating additional modalities like 3D structural data. This can be achieved by converting such modalities into sequence formats or integrating large models to collectively capture multi-modal information using fusion techniques. Moreover, combining Graph Neural Networks (GNNs) with transformers has led to significant advancements. For example, scMoFormer constructs cell-gene, gene-gene, protein-protein, and gene-protein graphs for multi-omics predictions, while DeepMAPS uses cell-gene graphs to estimate gene importance. GNNs excel in capturing local interactions, while transformers effectively model long-range dependencies, enabling comprehensive representations of intricate relationships in single-cell data. **Fourth,** novel tasks that can be explored in developing LLMs for bioinformatics include causal inference in multi-omics, such as determining how DNA variations influence mRNA abundance or protein expression. Spatial transcriptomics interpretation can model cell spatial organization within tissues. Epigenetic modulation prediction focuses on regulatory roles of histone modifications, DNA methylation, or chromatin accessibility. Synthetic biology applications can involve generating optimized gene or protein sequences, while cross-species genomics identifies conserved functional genomic elements. These tasks exemplify how LLMs can tackle emerging challenges in biological research. **Fifth,** developers should expand LLMs to accommodate emerging data types, such as CODEX imaging data and long-read sequencing data, which bring unique challenges in terms of data structure, preprocessing, and representation. **Lastly,** validation, application, and interpretability should be prioritized. Developers should not only evaluate models on specific tasks but also ensure that foundational challenges, such as the impact of sparsity in scRNA-seq data on cell type annotation performance, are fully addressed to enhance the robustness and utility of the models (**Figure 5**).

## Acknowledgements

We would like to express our gratitude to our colleagues and friends who provided invaluable advice and support throughout the duration of this study.

**Funding**

This work was partially supported by the National Institutes of Health [R01LM014156, R01GM153822, R01CA241930 to X.Z] and the National Science Foundation [2217515, 2326879 to X.Z]. The funders had no role in study design, data collection and analysis, decision to publish or preparation of the manuscript. Funding for open access charge: Dr & Mrs Carl V. Vartian Chair Professorship Funds to Dr. Zhou from the University of Texas Health Science Center at Houston.

*Conflict of interest statement.* None declared.

## Main figures

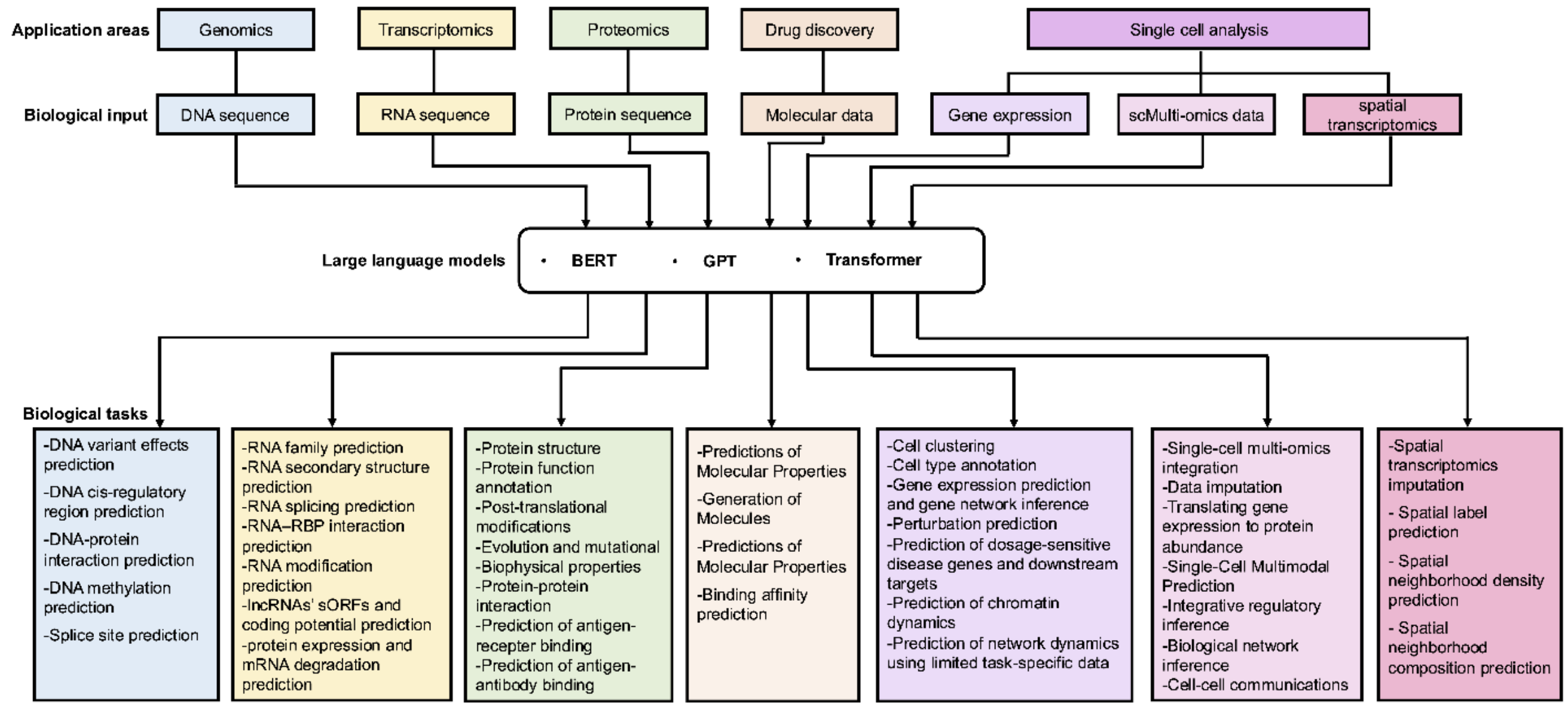

**Figure 1. Summary of the application of large language models in bioinformatics in this review.** Applications of large language models in bioinformatics include applications in genomics, transcriptomics, proteomics, drug discovery and single-cell analysis. Applications of LLMs in genomics focus on LLMs using DNA sequence; applications of LLMs in transcriptomics focus on using RNA sequence; applications of LLMs in proteomics focus on LLMs using protein sequence; applications of LLMs in drug discovery focus on LLMs using molecular data and applications of LLMs in single-cell analysis focus on LLMs using scRNA-seq, scMulti-omics and spatial transcriptomics data. Each corresponds to a variety of biological downstream tasks.

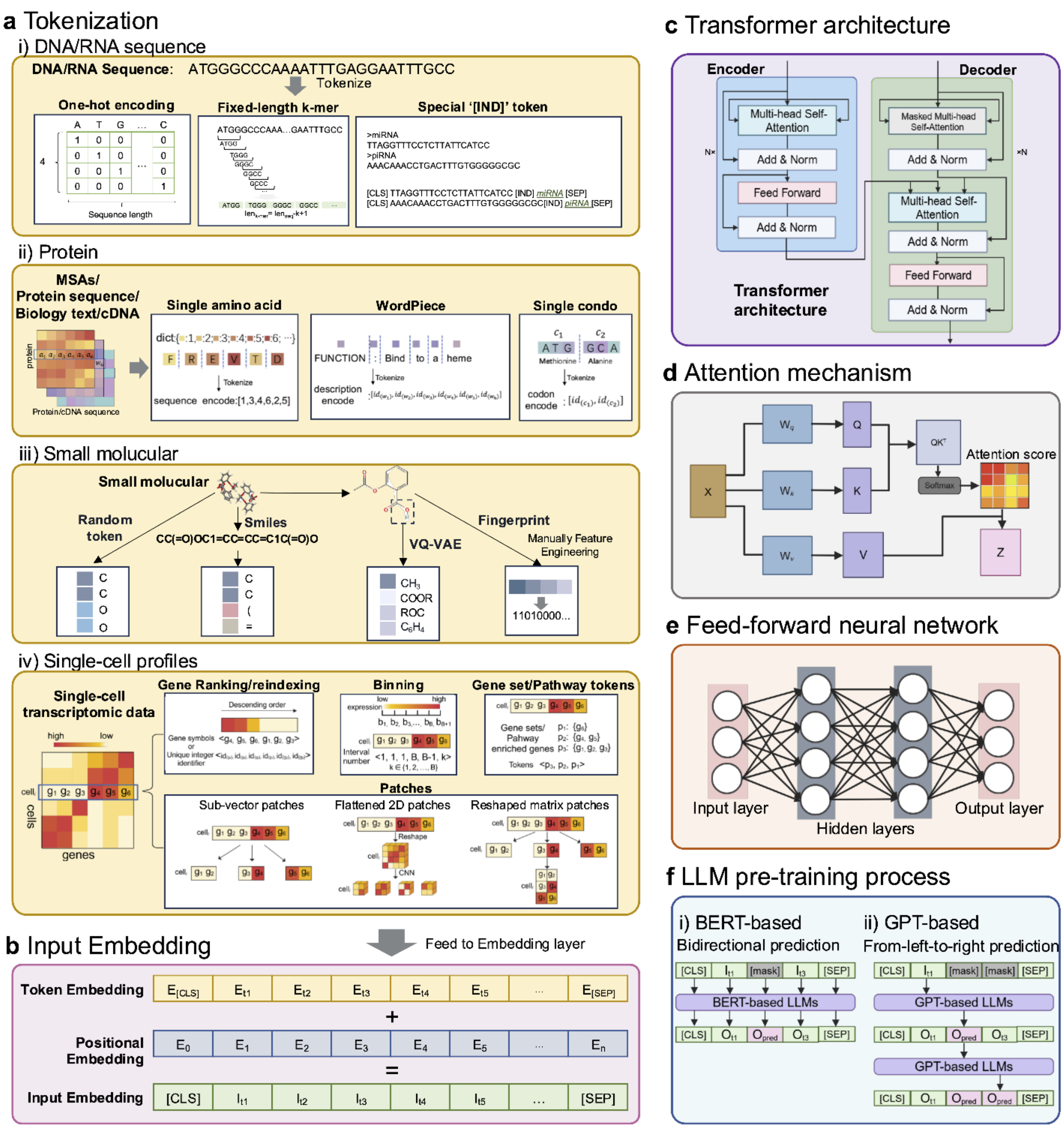

**Figure 2. Building blocks of large language models in bioinformatics. a,** tokenization methods tailored to various data types, including DNA/RNA sequences, proteins, small molecules, and single-cell data. **b,** input embedding strategies used in large language models to encode tokenized data. **c,** schematic representation of the transformer architecture, a foundational structure in LLMs. **d,** the attention mechanism, enabling models to focus on important features in sequences. **e,** the feed-forward network, a critical component of transformers for learning hierarchical

representations. **f,** pre-training processes for BERT and GPT-based models, highlighting BERT's bidirectional prediction approach and GPT's left-to-right prediction strategy.

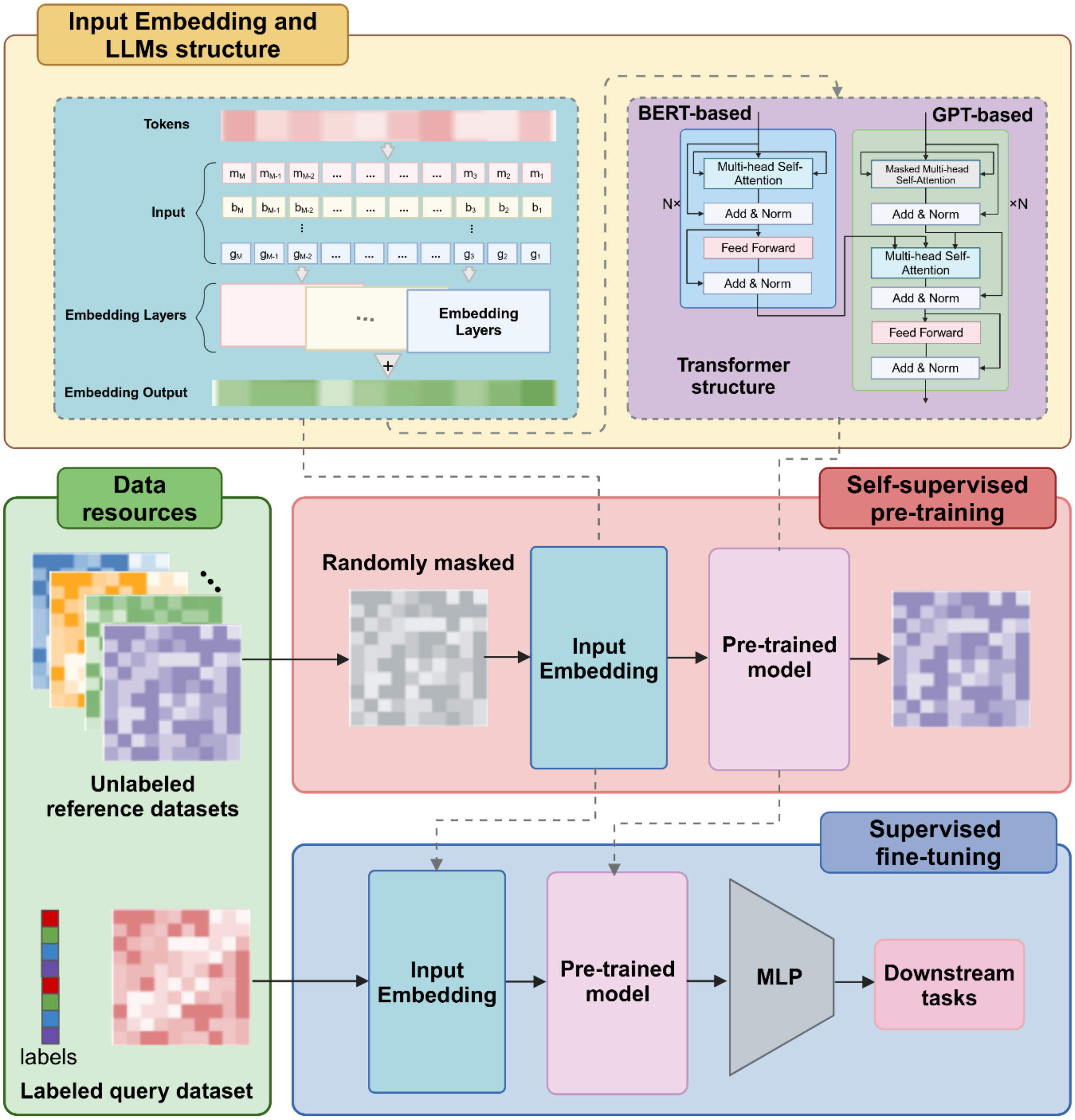

**Figure 3. Schematic diagram of the large language model pretraining and fine-tuning process.** The workflow begins with tokenizing the input data, which is then fed into the embedding layer and transformer models. The training process comprises two stages: pretraining and fine-tuning. Pretraining employs self-supervised learning on large-scale, unlabeled reference datasets to develop a general-purpose model with robust generalization capabilities. Fine-tuning builds upon the pretrained model, involving task-specific training to optimize performance for designated applications.

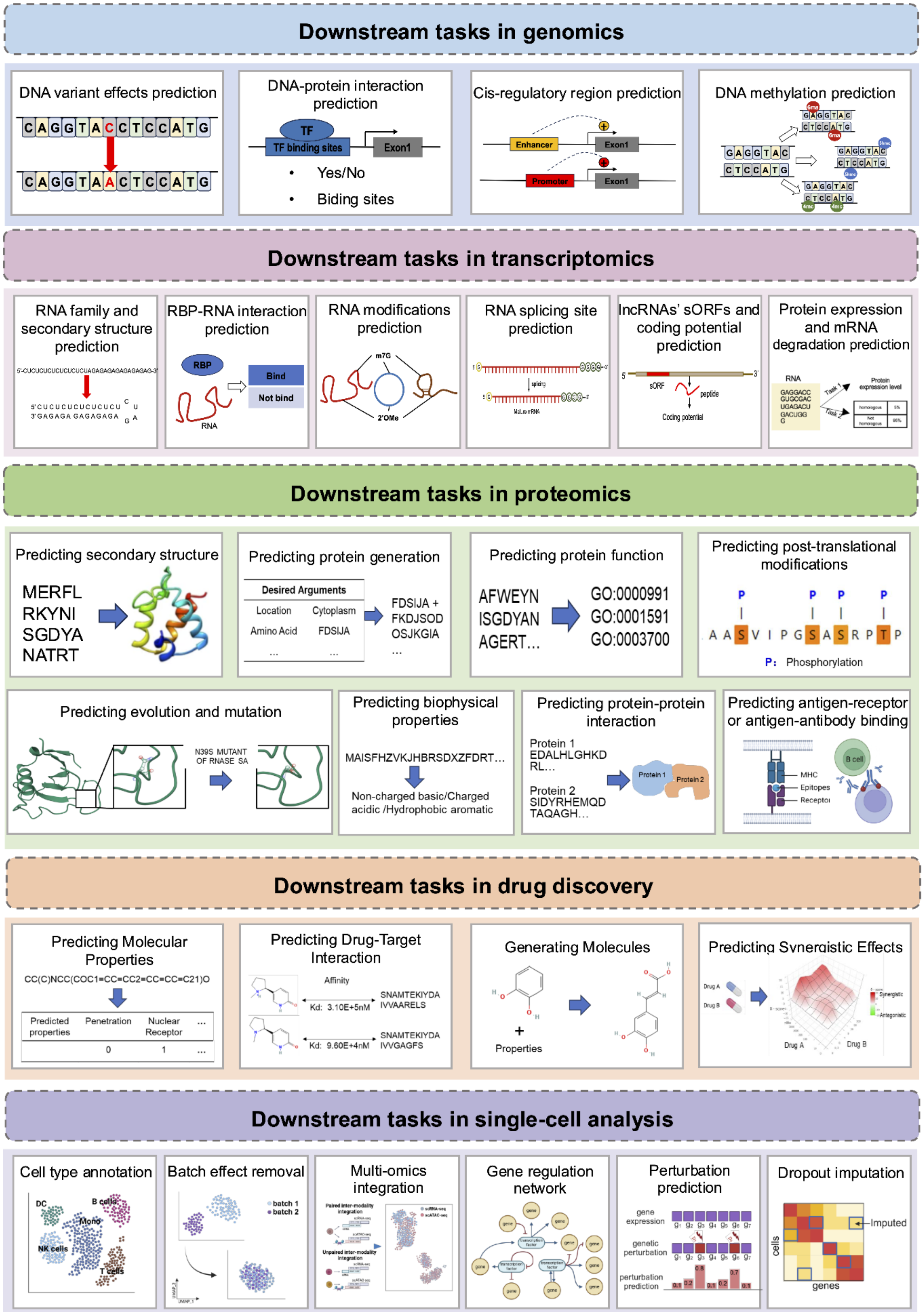
Downstream tasks in genomics
DNA variant effects prediction
CAGGTACCTCCATG
CAGGTAACTCCATG
DNA-protein interaction prediction
TF
TF binding sites
Exon1
Yes/No
Biding sites
Cis-regulatory region prediction
Enhancer
Promoter
Exon1
DNA methylation prediction
Downstream tasks in transcriptomics
RNA family and secondary structure prediction
RBP-RNA interaction prediction
RBP
Bind
Not bind
RNA
RNA modifications prediction
m7G
2'OMe
RNA splicing site prediction
lncRNAs' sORFs and coding potential prediction
sORF
peptide
Coding potential
Protein expression and mRNA degradation prediction
RNA
Downstream tasks in proteomics
Predicting secondary structure
MERFL
RKYNI
SGDYA
NATRT
Predicting protein generation
Desired Arguments
Location
Cytoplasm
Amino Acid
FDSIJA
FDSIJA +
FKDJSOD
OSJKGIA
Predicting protein function
AFWEYN
ISGDYAN
AGERT...
GO:0000991
GO:0001591
GO:0003700
Predicting post-translational modifications
A A S V I P G S A S R P T P
P: Phosphorylation
Predicting evolution and mutation
N39S MUTANT OF RNASE SA
Predicting biophysical properties
MAISFHZVKJHBRSDXZFDRT...
Non-charged basic/Charged acidic /Hydrophobic aromatic
Predicting protein-protein interaction
Protein 1
EDALHLGHKD
RL...
Protein 2
SIDYRHEMQD
TAQAGH...
Protein 1
Protein 2
Predicting antigen-receptor or antigen-antibody binding
B cell
MHC
Epitopes
Receptor
Downstream tasks in drug discovery
Predicting Molecular Properties
CC(C)NCC(COC1=CC=CC2=CC=CC=C21)O
Predicted properties
Penetration
Nuclear Receptor
0
1
Predicting Drug-Target Interaction
Affinity
Kd: 3.10E+5nM
SNAMTEKIYDA
IVVAARELS
Kd: 9.60E+4nM
SNAMTEKIYDA
IVVGAGFS
Generating Molecules
Properties
Predicting Synergistic Effects
Drug A
Drug B
Synergistic
Antagonistic
Downstream tasks in single-cell analysis
Cell type annotation
DC
B cells
Mono
NK cells
T cells
Batch effect removal
batch 1
batch 2
Multi-omics integration
Gene regulation network
Perturbation prediction
gene expression
genetic perturbation
perturbation prediction
Dropout imputation
Imputed
cells
genes

**Figure 4. Downstream tasks of large language models in bioinformatics.** Large language models (LLMs) have seen numerous successful applications in bioinformatics, addressing a wide array of tasks across DNA, RNA, protein, drug discovery, and single-cell analysis.

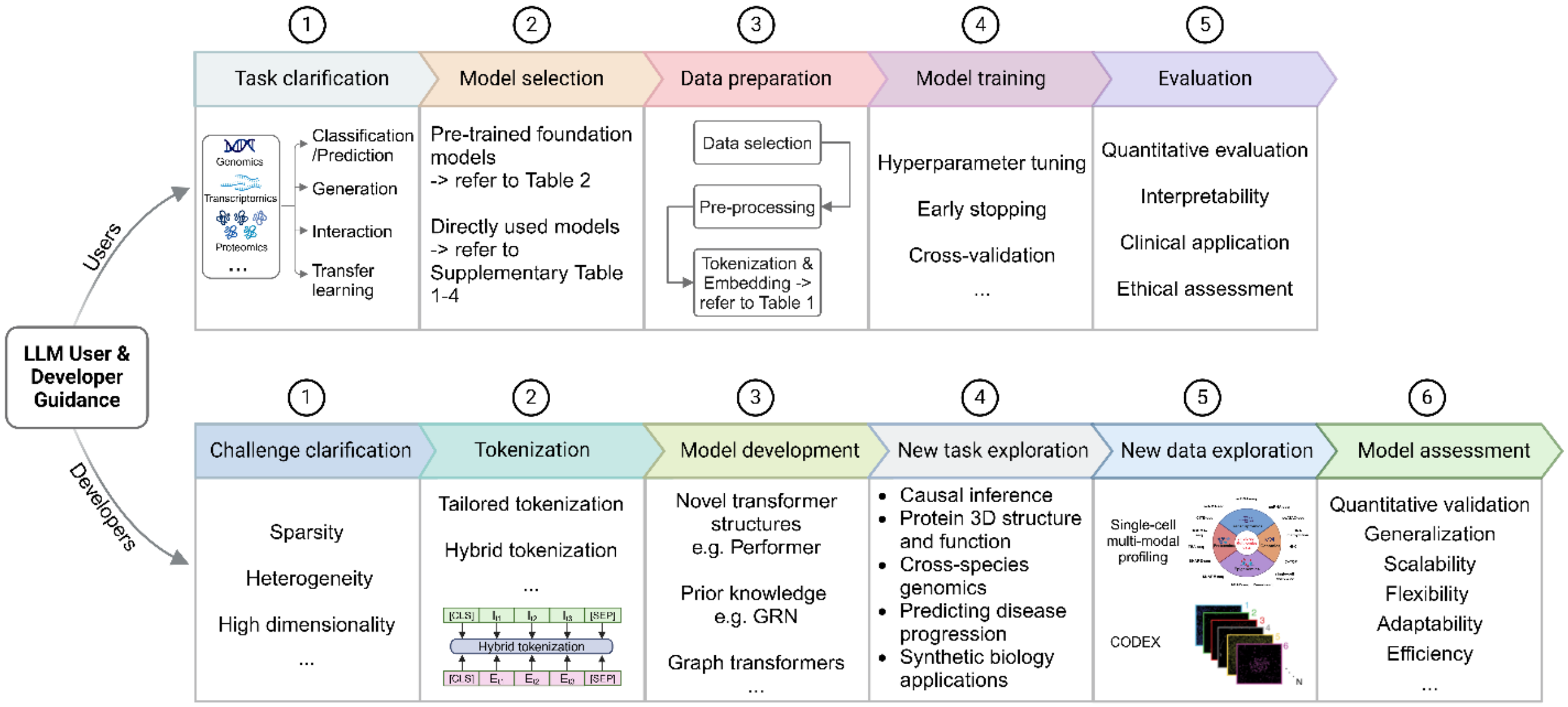

**Figure 5. Guidance for LLM users and developers on how to use and develop LLM in practice.** Guidance for LLM users includes steps such as clarifying the task, selecting an appropriate model, preparing the dataset, training the model, and evaluating its performance. For LLM developers, the focus involves identifying domain-specific challenges, designing tokenization strategies, advancing model architectures, exploring novel tasks and data types, and assessing model capabilities comprehensively.

## Tables

## Table 1. Tokenization methods for different types of data

| Application area | Data type | Method | Example |
|---|---|---|---|
| Genomics/Transcriptomics | DNA/RNA sequence | One-hot encoding | RNA-FM，RNA-MSM |
| | | Fixed-length k-mers | DNABERT，Nucleotide Transformer，DNABERT-2，DNAGPT，RNABERT |
| | | Special '[IND]' token | RNAErnie |
| Proteomics | MSAs/Protein sequences | Single Amino Acid Tokenization | MSA Transformer/TAPE, ESM-1b, ProtTrans, Progen |
| | Biomedical text | WordPiece | ProtST |
| | cDNA | Single condo Tokenization | CaLM |
| Drug discovery | Simplified Molecular-Input Line-Entry system (SMILES) | Random token | K-BERT |
| | | SmilesTokenizer | ChemBERTa, ChemBERTa-2, MolGPT |
| | | Graph VQ-VAE | Mole-BERT |
| | | fingerprint | SMILES-BERT |
| Single-cell analysis | Expression profiles | Gene expression Ranking | Geneformer, tGPT, iSEEEK |
| | | Binning | scBERT, scGPT, scFormer, CellLM, BioFormers, CancerFoundation |
| | | Gene set/Pathway tokens | TOSICA |
| | | Patches | CIForm, scTranSort, scCLIP |
| | | Gene value projection | scTranslator, scFounfation, scMulan, scGREAT |
| | | Cell tokens | CellPLM, ScRAT, mcBERT |

**Table 2. Foundation models in bioinformatics**

| Application area | Model | Architecture | Pre-training Data | Code available |
|---|---|---|---|---|
| Genomics | GPN | Transformer-based | Reference genomes from 8 species | *https://github.com/songlab-cal/gpn* |
| | Nucleotide Transformer | Transformer-based | 3.2 billion nucleotides in GRCh38/hg38 reference assembly, 20.5 trillion nucleotides including 125 million mutations (111 million SNPs, 14 million indels), and 174 billion nucleotides from 850 species | *https://github.com/instadeepai/nucleotide-transformer* |
| | DNABERT | BERT-based | 2.75 billion nucleotide based human genome dataset | *https://github.com/jerryji1993/DNABERT* |
| | DNABERT-2 | BERT-based | 2.75 billion nucleotide based human genome dataset and 32.49 billion nucleotide bases from 135 species, spread across 6 categories | *https://github.com/MAGICS-LAB/DNABERT_2* |
| | MoDNA | BERT-based | Same as Nucleotide Transformer | *https://github.com/uta-smile/MoDNA* |
| | GROVER | BERT-based | Homo sapiens (human) genome assembly GRCh37 (hg19) | *https://github.com/rowanz/grover* |
| | MuLan-Methyl | BERT-based | 3 main types of DNA methylation sites (6mA, 4mC, and 5hmC) across 12 genomes, in total 250,599 positive samples | *https://github.com/husonlab/mulan-methyl* |
| | iDNA-ABF | BERT-based | Same as MuLan-Methyl | *https://github.com/FakeEnd/iDNA_ABF* |
| | iDNA-ABT | BERT-based | Same as MuLan-Methyl | *https://github.com/YUYING07/iDNA_ABT* |
| | DNAGPT | GPT-based | Reference genomes from the Ensembl database include 3 billion bps, with a total of 1,594,129,992 bps across 9 species | *https://github.com/TencentAILabHealthcare/DNAGPT* |
| Transcriptomics | RNABERT | BERT-based | 76 237 human-derived small ncRNAs from RNAcentral | *https://github.com/mana438/RNABERT* |
| | RNA-FM | BERT-based | About 27 million ncRNA sequences across 47 different databases | *https://github.com/ml4bio/RNA-FM* |
| | RNA-MSM | BERT-based | 4069 RNA families from rfam | *https://github.com/yikunpku/RNA-MSM* |
| | SpliceBERT | BERT-based | 2 million sequences and approximately covering 65 billion nucleotides of 72 vertebrates from UCSC genome browser | *https://github.com/biomed-AI/SpliceBERT* |
| | UNI-RNA | BERT-based | 23 million ncRNA sequences obtained from the RNAcentral database | *https://github.com/ComDec/unirna-tools* |
| | 3UTRBERT | BERT-based | 108,573 unique mRNA transcripts from the GENCODE and each contains 3,754 nucleotides (median 3048 nts) on average. | *https://github.com/yangyn533/3UTRBERT* |
| | UTR-LM | BERT-based | 214,349 unlabeled 5′ UTR sequences from Ensembl across 5 species | *https://github.com/a96123155/UTR-LM* |
| | RNAErnie | Transformer- based | 23 million ncRNA sequences obtained from the RNAcentral database | *https://github.com/CatIIIIIIII/RNAErnie* |
| Proteomics | TAPE | Transformer-based | 31 million protein sequences from Pfam | *https://github.com/songlab-cal/tape* |
| | ESM-1b | Transformer-based | 250 million protein sequences from UniRef50 | *https://github.com/facebookresearch/esm* |
| | ProtTrans | Transformer-XL, XLNet, BERT, Albert, Electra, T5 | About 2.3 billion protein sequences from UniRef and BFD | *https://github.com/agemagician/ProtTrans* |
| | ProtGPT2 | GPT-based | 50 million protein sequences from UniRef50 | *https://huggingface.co/docs/transformers/main_classes/trainer* |
| | ProteinBERT | BERT-based | 106 million protein sequences with GO annotations from UniRef50 | *https://github.com/nadavbra/protein_bert* |
| | KeAP | BERT-based | 5 million Triplet in the format of (Protein, Relation, Attribute) with nearly 600k protein, 50k attribute terms, and 31 relation terms included | *https://github.com/RL4M/KeAP* |
| | CaLM | Transformer-based | 9,858,385 cDNA sequences of seven model organisms | *https://github.com/oxpig/CaLM* |
| Drug discovery | SMILES-BERT | BERT-based | Two datasets from NCATS (NIH) and 128 datasets from PubChem | *https://github.com/uta-smile/SMILES-BERT* |
| | ChemBERTa | BERT-based | 77 million unique SMILES | *https://github.com/seyonechithrananda/bert-loves-chemistry* |
| | K-BERT | BERT-based | Book review dataset contains 20,000 positive and 20,000 negative reviews collected from Douban | *https://github.com/autoliuweijie/K-BERT* |
| | Mole-BERT | BERT-based | 2 million molecules | *https://github.com/junxia97/Mole-BERT* |
| | MolGPT | GPT-based | Datasets from MOSES and GuacaMol | *https://github.com/devalab/molgpt* |
| | ProtBERT | BERT-based | Datasets from Uniref50, UniRef100 and BFD | *https://github.com/agemagician/ProtTrans/* |
| | DeepDDS | BERT-based | Datasets from NCI-ALMANAC | *https://github.com/sorachel/DFFNDDS* |
| | SynerGPT | GPT-based | Datasets from DrugCombDB | *Code will be made available upon publication* |
| Single-cell analysis | scBERT | BERT-based | 1,126,580 cells from 209 datasets across 74 tissues and 451,513 cells from four sequencing platforms | *https://github.com/TencentAILabHealthcare/scBERT* |
| | scGPT | GPT-based | 33 million human cells from the CellXGene collection | *https://github.com/bowang-lab/scGPT* |
| | Geneformer | BERT-based | 29.9 million human single-cell transcriptomes | *https://huggingface.co/ctheodoris/Geneformer* |
| | scFoundation | BERT-based | About 50 million human single-cell transcriptomic profiles | *https://github.com/biomap-research/scFoundation* |
| | tGPT | GPT-based | 22.3 million single-cell transcriptomes | *https://github.com/deeplearningplus/tGPT* |
| | GeneCompass | BERT-based | over 120 million single-cell transcriptomes from humans and mice | *https://github.com/xCompass-AI/GeneCompass* |
| | scMulan | GPT-based | More than 10 million manually annotated single-cell RNA-seq data | *https://github.com/SuperBianC/scMulan* |
| | UCE | BERT-based | 300 datasets from the CellXGene corpus includes over 36 million cells, 1,000+ cell types, dozens of tissues, and eight species | *https://github.com/snap-stanford/uce* |
| | scPRINT | BERT-based | More than 50M cells from theCellXGene database | *https://github.com/cantinilab/scPRINT* |
| | CancerFoundation | BERT-based | 50 million cells with roughly a quarter being tumor cells | *https://github.com/BoevaLab/CancerFoundation* |
| | Nicheformer | BERT-based | 57 million dissociated and 53 million spatially resolved cells across 73 tissues from both human and mouse | *https://github.com/theislab/nicheformer* |

**Table 3. Large language models for downstream tasks in bioinformatics**

| Input data | Biological tasks | Models |
|---|---|---|
| DNA sequence | Genome-wide variant effects prediction | DNABERT, DNABERT-2, GPN, Nucleotide Transformer |
| | DNA cis-regulatory regions prediction | DNABERT, DNABERT-2, BERT-Promoter, iEnhancer-BERT, Nucleotide Transformer |
| | DNA-protein interaction prediction | DNABERT, DNABERT-2, TFBert, GROVER, and MoDNA |
| | DNA methylation (6mA,4mC 5hmC) prediction | BERT6mA, iDNA-ABF, iDNA-ABT, and MuLan-Methyl |
| | RNA splice sites prediction from DNA sequence | DNABERT, DNABERT-2 |
| RNA sequence | RNA 2D/3D structure prediction | RNA-FM, RNA-MSM, and RNA-FM |
| | RNA structural alignment, RNA family clustering | RNABERT |
| | RNA splice sites prediction from RNA sequence | SpliceBERT |
| | RNA N7-Methylguanosine modification prediction | BERT-m7G |
| | RNA 2'-O-methylation Modifications prediction | Bert2Ome |
| | Multiple types of RNA modifications prediction | Rm-LR |
| | Predicting the association between miRNA, lncRNA and disease | BertNDA |
| | Identifying lncRNAs | LncCat |
| | Protein expression and mRNA degradation prediction | CodonBERT |
| Protein sequences<br>MSAs<br>Gene ontology annotations<br>Triplets of protein-relation-attribute<br>Protein property descriptions<br>cDNA sequences | Secondary structure and contact prediction | MSA Transformer, ProtTrans, SPRoBERTa, TAPE, KeAP |
| | Protein sequence generation | ProGen, ProtGPT2 |
| | Protein function prediction | SPRoBERTa, ProtST, PromptProtein, CaLM |
| | Major PTMs prediction | ProteinBERT |
| | Evolution and mutation prediction | SPRoBERTa, UniRep, ESM-1b, TAPE, PLMsearch, DHR |
| | Biophysical properties prediction | TAPE, PromptProtein |
| | Protein-protein interaction and binding affinity prediction | KeAP |
| | Antigen-Receptor binding prediction | MHCRoBERTa, BERTMHC, TCR-BERT, SC-AIR-BERT, Antiformer |
| | Antigen-Antibody binding prediction | AbLang, AntiBERTa, EATLM |
| Molecular SMILES | Predicting Molecular Properties | SMILES-BERT, ChemBERTa, K-BERT |
| | Generating Molecules | MolGPT |
| Molecular graphs | Predicting Molecular Properties | MOLE-BERT |
| Molecular fingerprints and protein sequences | Predicting Drug-Target Interaction | TransDTI, FG-BERT |
| Molecular SMILES and protein sequences | Predicting Synergistic Effects | SynerGPT, C2P2 |
| scRNA-seq data | Cell clustering | tGPT, scFoundation, UCE, iSEEEK, CellPLM, BioFormers, mcBERT |
| | Cell type annotation | scBERT, scGPT, CIForm, TOSICA, scTransSort, TransCluster, Geneformer, GeneCompass, scMulan, CellLM, CellPLM, scPRINT |
| | New cell type identification | scBERT, TOSICA, UCE |
| | Batch effect removal | scBERT, scGPT, CIForm, TOSICA, Geneformer, scMulan, iSEEEK, scPRINT, CancerFoundation, mcBERT |
| | Trajectory inference/Pseudotime analysis | tGPT, scMVP, iSEEEK |
| | Drug response/sensitivity prediction | scFoundation, CellLM, CancerFoundation |
| | Gene network inference | scGPT, Geneformer, GeneCompass, iSEEEK, scGREAT, BioFormers, scPRINT |
| | Gene perturbation prediction | scGPT, scFoundation, GeneCompass, CellPLM, BioFormers |
| | Gene expression prediction | scGPT, scMVP, scFoundation, GeneCompass, CellPLM, BioFormers |
| | cis-regulatory element identification | scMVP |
| | Drug dose-response prediction, Gene dosage sensitivity prediction | GeneCompass |
| scMuti-omics data | Single-cell multi-omics integration | scGPT, scMVP, DeepMAPS, scCLIP |
| | Biological network inference<br>Cell-cell communications | DeepMAPS |
| | Translating gene expression to protein abundance | scTranslator, scMoFormer |
| | Single-cell multimodal prediction | scMoFormer |
| | Integrative regulatory inference | scTranslator |
| Single-cell spatial transcriptomics data | Spatial transcriptomics imputation | CellPLM, Nicheformer, SpaFormer |
| | Spatial label prediction<br>Spatial neighborhood density prediction<br>Spatial neighborhood composition prediction | Nicheformer |

## Supplementary figures

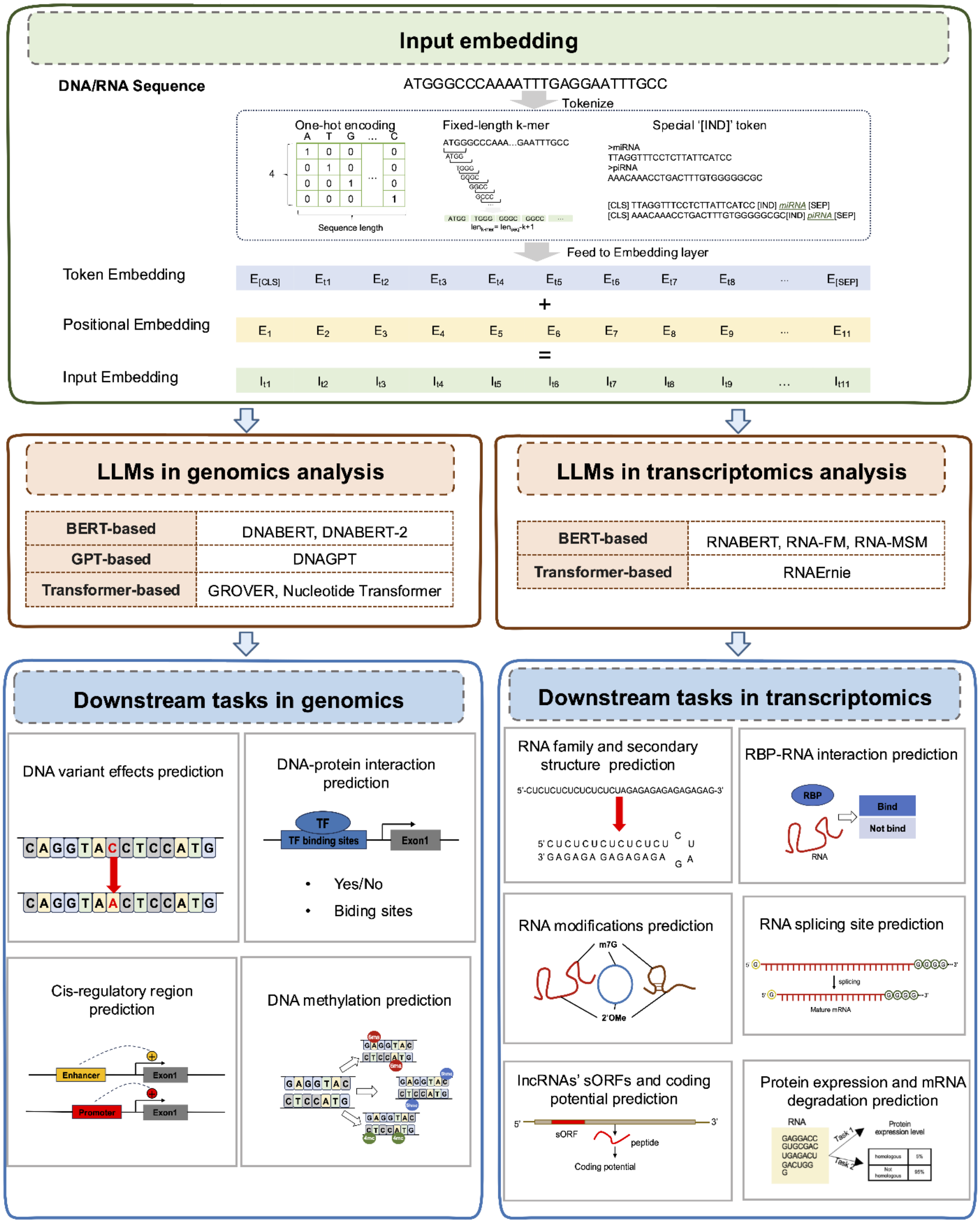

**Supplementary figure 1. Applications of large language models in genomics and transcriptomics.** DNA language models take DNA sequence as input, use transformer, BERT,

GPT models to solve multiple biological tasks, including genome-wide variant effects prediction, DNA cis-regulatory regions prediction, DNA-protein interaction prediction, DNA methylation (6mA,4mC 5hmC) prediction, splice sites prediction from DNA sequence. The RNA language models take RNA sequences as input, use transformer, BERT, GPT models to solve multiple biological tasks, including RNA 2D/3D structure prediction, RNA structural alignment,, RNA family clustering, RNA splice sites prediction from RNA sequence, RNA N7-methylguanosine modification prediction, RNA 2'-O-methylation modifications prediction, multiple types of RNA modifications prediction, predicting the association between miRNA, lncRNA and disease, identifying lncRNAs, lncRNAs' coding potential prediction, protein expression and mRNA degradation prediction.

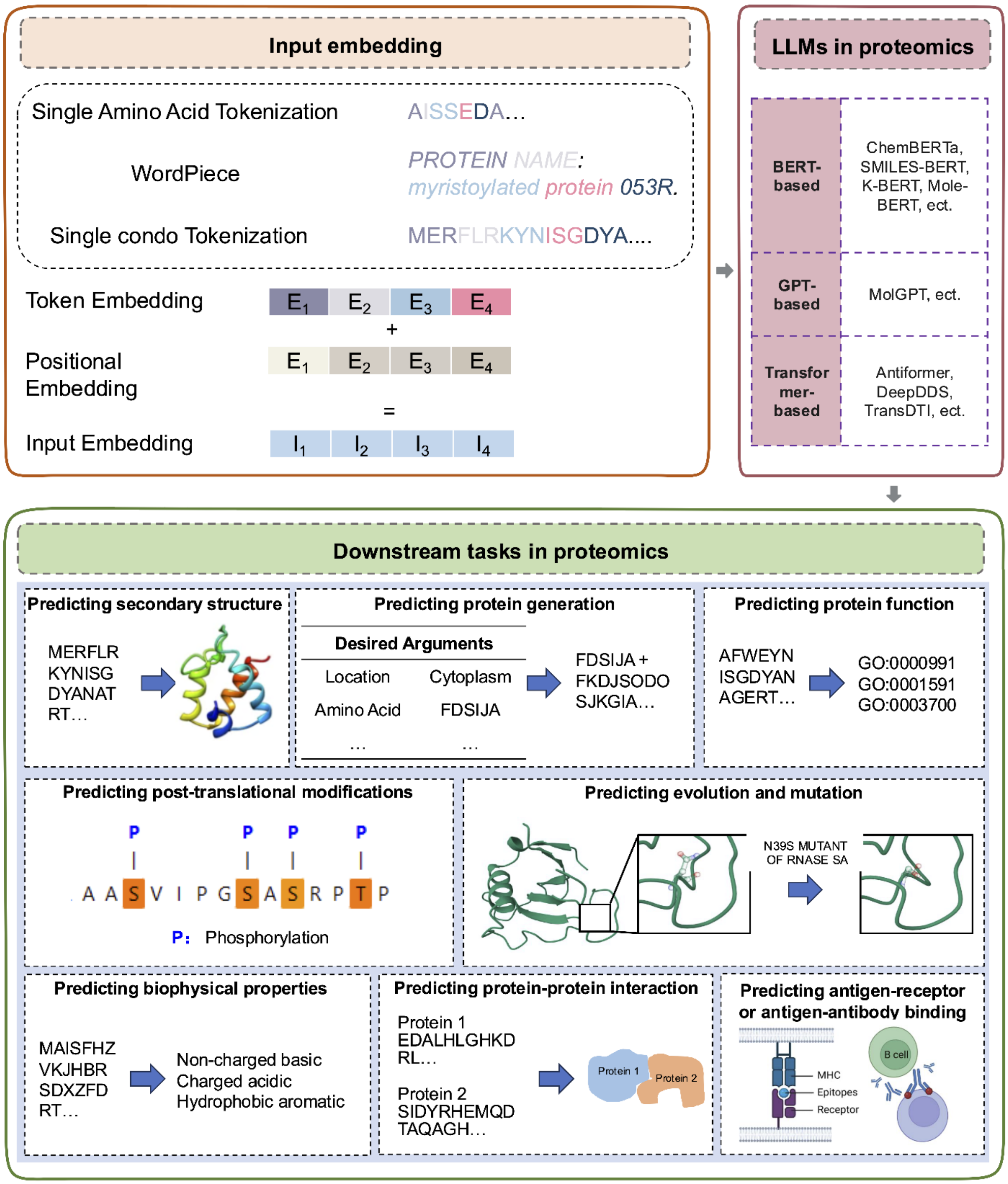

**Supplementary figure 2. Applications of large language models in proteomics.** The protein language models take multiple sequence alignment, protein sequence, gene ontology and protein-relation-attribute as input, use transformer, BERT, GPT models to solve multiple biological tasks, including predicting secondary structure, predicting protein generation, predicting protein function,

predicting post-translational modifications, predicting evolution and mutation, predicting biophysical properties, predicting protein-protein interaction and predicting antigen-receptor or antigen-antibody binding.

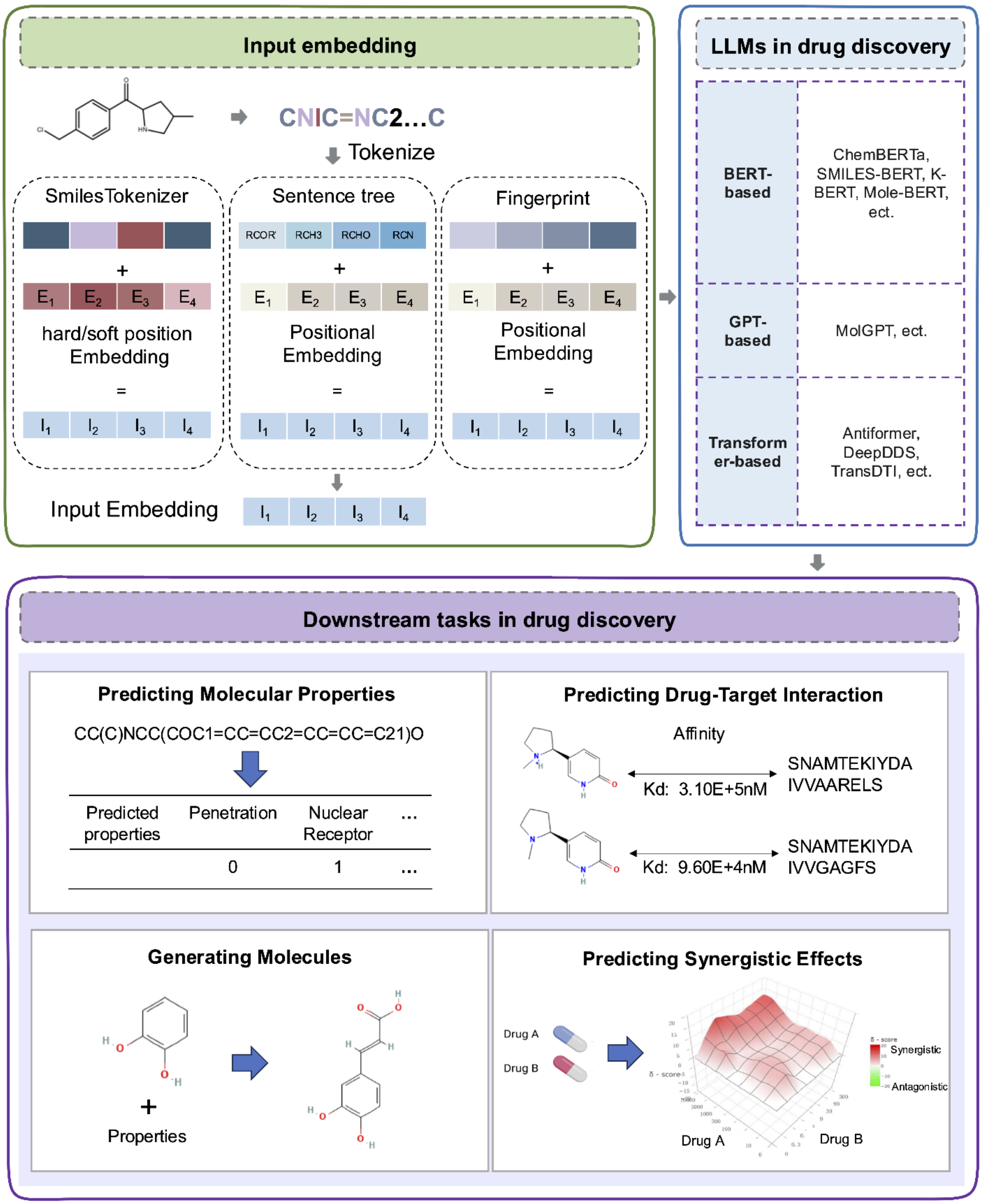

**Supplementary figure 3. Applications of large language models in drug discovery.** The language models for drug discovery take molecular SMILES, protein sequence, molecular fingerprints and molecular graphs as input, use transformer, BERT, GPT models to solve multiple

biological tasks, including predicting molecular properties, predicting drug-target interaction, generating molecules and predicting synergistic effects.

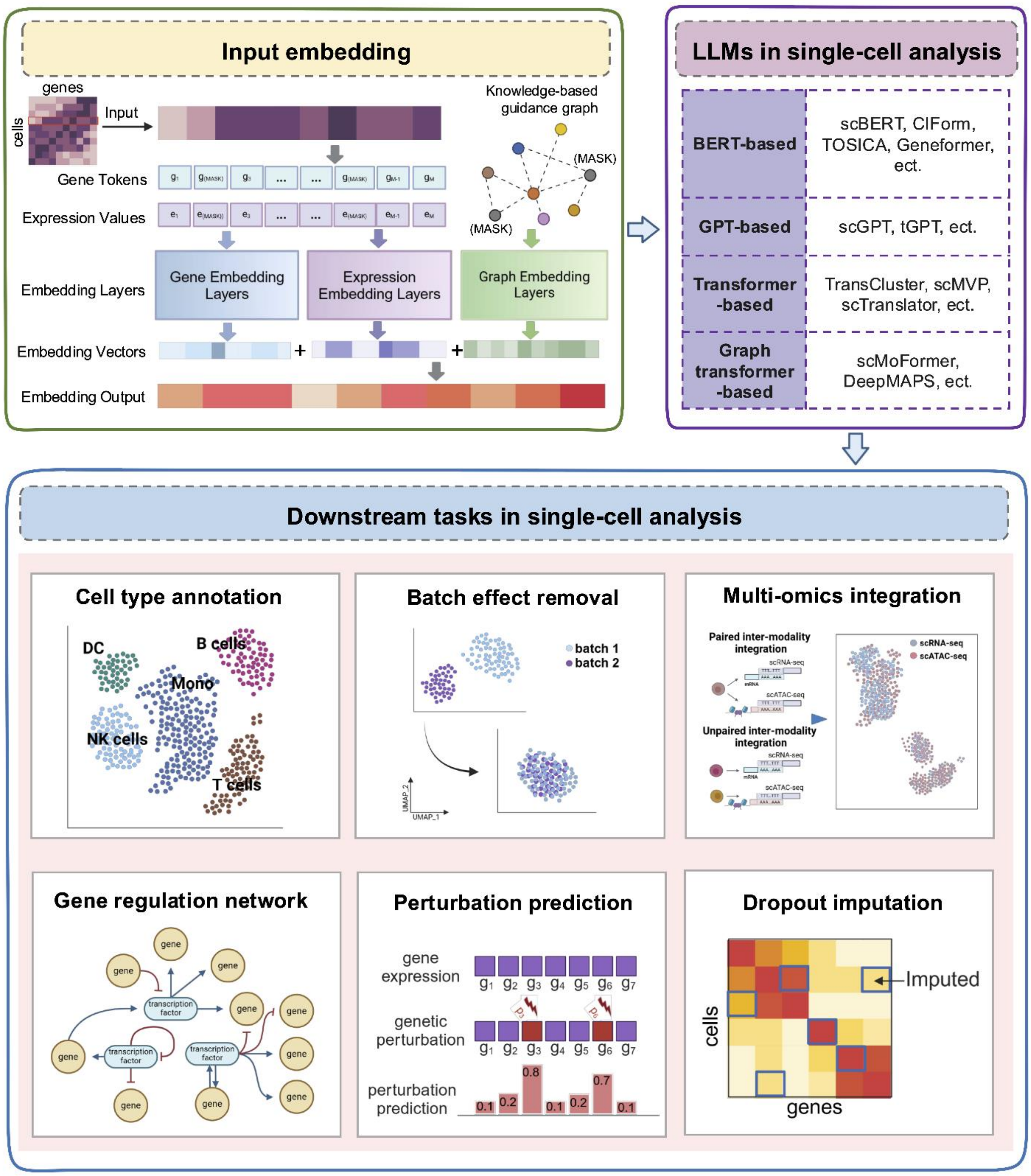

**Supplementary figure 4. Applications of large language models in single-cell analysis.** The single-cell language models take gene expression or single-cell multi-omics data as input, use transformer, BERT, GPT models to solve multiple biological tasks, including cell type annotation, batch effect removal, multi-omics integration, gene regulation network inference perturbation prediction, dropout imputation.

## Supplementary Tables

## Supplementary Table 1. Detailed information of large language models for genomic and transcriptomic tasks

| Application area | Models | Ref | Publication time | Parameters | Architecture | Fine-tuning datasets | | | Downstream tasks |
|---|---|---|---|---|---|---|---|---|---|
| | | | | | | **Data type** | **Source** | **Size** | |
| DNA sequence language model | DNABERT | [1] | Aug 2021 | 12 transformer layers with 768 hidden units and 12 attention heads in each layer | BERT-based | DNA sequence | Human TATA and non-TATA promoters of 10 000 bp length[2] and ChIP-seq dataset [3] | 3,065 human TATA and 26,533 non-TATA promoter-containing sequences and 690 ChIP-seq dataset covers 161 transcription factor binding profiles in 91 human cell lines | Transcription factor binding sites prediction |
| | | | | | | | DNA sequence | - | Motif analysis |
| | | | | | | | Assembly GRCh38 FASTA file [4] | 10,000 donor, acceptor, and non-splice site sequences | Splice donor and acceptor sites prediction |
| | | | | | | | dbSNP release 153 [5] | 700 million short genetic variants | Identifying effects of genetic variants |
| | DNABERT-2 | [6] | July 2023 | batch size is 32, warmup step is 50, and weight decay is 0.01 | BERT-based | DNA sequence | TATA and non-TATA promoters downloaded from Eukaryotic Promoter Database (EPDnew) [2] | 3,065 human TATA and 26,533 non-TATA promoter-containing sequences | Promoter detection and core promoter detection |
| | | | | | | | ChIP-seq datasets [7] | 161 TF binding profiles in 91 human cell lines(human) and 78 mouse ENCODE ChIP-seq data | Transcription factor binding site prediction |
| | | | | | | | Ensembl GRCh38 human reference genome [4] | 10,000 splice donors, acceptors, and non-splice site sequences. | Splice site prediction |
| | | | | | | | Histone modification (Yeast) | H3, H3K14ac, H3K36me3, H3K4me1, H3K4me2, H3K4me3, H3K79me3, H3K9ac, H4, H4ac | Epigenetic marks prediction |
| | | | | | | | SARS_CoV_2 variants [8] | 9 types of SARS_CoV_2 variants, including *Alpha, Beta,Delta, Eta, Gamma, Iota, Kappa, Lambda and Zeta.* | Covid variant prediction |
| | Nucleotide | [9] | Jan 2023 | 2 hidden layers | Transfor | DNA | Annotated DNA sequence | 90,000 sequences annotated by | Detect known genomic |
| | Transformer | | | | mer-based | sequence | [10] | Ensembl (“5’ UTR”, “3’ UTR”, “exon”, “intron”, “enhancer”, “promoter”, “CTCF binding site”, “open chromatin”, and “transcription factor binding sites”. ) | elements |
| | | | | | | | DNA sequence with SNP [11] | Independent dataset of genetically diverse human genomes, originating from 7 different meta-populations | Detect human genetic variation |
| | | | | | | | 1000 Genomes Project SNPs [12] | chromosome 22 sequence with 17 variant categories (e.g. stop gained, missense, intergenic) | Predict the impact of mutations |
| | DNAGPT | [13] | July 2023 | 12 layers of transformer blocks based on unidirectional attention, with each layer containing 12 attention heads and a hidden layer size of 768 | GPT-based | DNA sequence | DNA sequence from DeepGSR [14] | 20,933, 18,693, 12,082, and 27,203 true polyadenylation signals data; and 28,244, 25,205, 17,558, and 30,283 true translation initiation sites for human, mouse, bovine, and fruit fly, respectively which are used as ground-truth, non-genomic signals and regions sequences from the genome sequences and combined them with the true cases | Genomic signals and regions prediction |
| | | | | | | | DNA sequence from Xpresso [15] | 18,377 and 21,856 promoters as well as the mRNA half-lives in human and mouse respectively and held out 1000 cases in each specie | mRNA expression level prediction |
| | GROVER | [16] | July 2023 | 12 transformer layers, 5,000 embeddings | BERT-based | DNA sequence | CTCF ChIP-seq data [17] | ~85,000 binding motifs, only ~32,000 are indeed bound by CTCF | Protein-DNA binding prediction |
| | GPN | [18] | BioRxiv posted April 2023 | 25 convolutional blocks with a feed-forward layer, 512embedding sizes of the pre-trianed foundation model | BERT-based | DNA sequence | DNA sequence | - | DNA motifs predictions |
| | | | | | | | 1001 Genomes Project [12] | 10 million SNPs | Variant effect prediction |
| | BERT-Promoter | [19] | Aug 2022 | BERT model included 12 layers, 768-hidden, 12 heads, and 110,000,000 parameters | BERT-based | DNA sequence | ChIP-chip data, gSELEX peaks, ChIP-exo plus RNA-seq [20, 21] | 3382 promoters (1591 strong promoter samples and 1791 weak promoter samples) and 3382 non-promoters | DNA promoter prediction |
| | TFBert | [22] | Mar 2023 | 12-layer encoder | BERT-based | DNA sequence | ChIP-seq datasets [7] | 690 ChIP-seq dataset contains a training set (80%) and a corresponding test set (20%) | DNA–protein binding sites prediction |
| | MoDNA | [23] | Aug 2022 | - | BERT-based | DNA sequence | Same experiment data with DNABERT [2, 3] | 3,065 human TATA and 26,533 non-TATA promoter-containing sequences and 690 ChIP-seq dataset covers 161 transcription factor binding profiles in 91 human cell lines | Promoter Prediction |
| | | | | | | | CHIP-Seq datasets [3] | 690 CHIP-Seq datasets of uniform TFBS contains 161 TFs covering 91 human cell types | Transcription Factor Binding Sites Prediction |
| | iEnhancer-BERT | [24] | Aug 2022 | 12-layer transformer architecture with simple fine-tuning | BERT-based | DNA sequence | 15 chromatin states of 9 cell types [25] | 2968 samples including 1484 non-enhancers, 742 strong enhancers and 742 weak enhancers | Identifying Enhancers and Their Strength |
| | BERT6mA | [26] | Mar 2022 | The hidden size of the LSTM unit is set to 128.DNA sequence embedding with Word2vec. | BERT-based | DNA sequence | Nuclei purification, MNase-seq and ChIP-seq [27-31] | 6mA and non6mA data in 11 species including Arabidopsis thaliana (31873 6mAs and non-6mAs), Caenorhabditis elegans (79616 mAs and non-6mAs), Casuarina equisetifpolia (6066 6mAs and non-6mAs), Drosophila melanogaster (11 191 6mAs and non-6mAs), Fragaria vesca (3102 6mAs and non-6mAs), H. sapiens (18 335 6mAs and non-6mAs), Rosa chinensis (599 6mAs and non-6mAs), Saccharomyces cerevisiae (37866mAs and non-6mAs), Thermus thermophilus (107 600 6mAs and non-6mAs), Ts. SUP5–1 (3379 | DNA N6-methyladenine site prediction |
| | | | | | | | | 6mAs and non6mAs) and Xoc. BLS256 (17 215 6mAs and non-6mAs) | |
| | iDNA-ABF | [32] | Oct 2022 | 12 transformer layers with 768 hidden units and 12 attention heads in each layer | BERT-based | DNA sequence | ChIP-seq data, ATAC-seq data, and histone modifications (HM) data of three human cell lines [33, 34], and DNA methylation dataset from the iDNA-MS [35] | 3 main types of DNA methylation sites (6mA,4mC, and 5hmC) across 12 genomes (1 bacteria and 11 eukaryotes), in total 250,599 positive samples | DNA 6mA, 4mC, and 5hmC prediction |
| | | | | | | | DNA sequence | - | Motifs analysis |
| | iDNA-ABT | [36] | Sep 2021 | 12 transformer layers with 12 attention heads in each layer. | BERT-based | DNA sequence | ChIP-seq data, ATAC-seq data, and histone modifications (HM) data of three human cell lines [33, 34] , and DNA methylation dataset from the iDNA-MS [35] | 3 main types of DNA methylation sites (6mA,4mC, and 5hmC) across 12 genomes (1 bacteria and 11 eukaryotes), in total 250,599 positive samples | DNA 6mA, 4mC, and 5hmC prediction |
| | | | | | | | DNA sequence | - | Motifs analysis |
| | MuLan-Methyl | [37] | July 2023 | 12 layers in the encoder stack, 768 hidden units for feed-forward networks, and 12 attention heads. | BERT-based | DNA sequence | ChIP-seq data, ATAC-seq data, and histone modifications (HM) data of three human cell lines [33, 34] , and DNA methylation dataset from the iDNA-MS [35] | 3 main types of DNA methylation sites (6mA,4mC, and 5hmC) across 12 genomes (1 bacteria and 11 eukaryotes), in total 250,599 positive samples | DNA 6mA, 4mC, and 5hmC prediction |
| RNA sequence language model | RNA-MSM | [38] | Nov 2023 | 12 attention heads with embedding size of 768 | BERT-based | RNA sequence | RNA secondary structure and three-dimensional RNA structures [39] | The training, validation, and test sets have 405, 40, and 70 RNAs. | RNA secondary structure prediction |
| | | | | | | | RNA secondary structure and three-dimensional RNA structures [39] | The training, validation, and test sets have 405, 40, and 70 RNAs. | RNA solvent accessibility prediction |
| | RNA-FM | [40] | Arxiv posted Apr 2022 | 12 transformer-based bidirectional encoder blocks and 640 embedding | BERT-based | RNA sequence | RNA secondary structure [41, 42] | 37149 structures from 8 RNA types of RNAStralign and 3975 RNA structures from 10 RNA types of ArchiveII | RNA secondary structure prediction |
| | | | | | | | RNA secondary structure [41, 42] | 37149 structures from 8 RNA types of RNAStralign and 3975 RNA structures from 10 RNA types of ArchiveII | RNA 3D closeness prediction |
| | | | | | | | whole genome of Severe acute respiratory syndrome coronavirus 2 (SARS-CoV-2) [43] | Whole genome | SARS-CoV-2 genome structure and evolution prediction |
| | | | | | | | In vivo RNA secondary structure profiles for RNA-protein interaction [44] | - | Protein-RNA interaction prediction |
| | | | | | | | Human 5'UTR library [45] | 83,919 5'UTRs of 75 different lengths and their corresponding mean ribosome loadings | mRNA 5' UTR-based mean ribosome loading prediction |
| | RNABERT | [46] | Feb 2022 | 6 hidden layers of BERT, an embedding layer, one bidrectional-LSTM unit, two dense layers one with ReLU activation and a softmax output layer of LSTM | BERT-based | RNA sequence | RNA (ncRNA) families from RFam database[46] | 31 RNA families | classifying RNA families RNA secondary structure prediction |
| | SpliceBERT | [47] | Mar 2024 | 6 transformer encoder layers, 512 hidden layer and 16 attention heads | BERT-based | RNA sequence | Reference genomes in fasta [48] | The pre-mRNA sequences from 72 vertebrate genomes for pre-training | Estimating splice sites |
| | BERT-m7G | [49] | Aug 2021 | - | BERT-based | RNA sequence | RNA sequence with N7-methylguanosine sites and RNA sequence without N7-methylguanosine sites AlkAniline-Seq, MeRIP-seq, and miCLIP-seq [50] | 741 RNA sequences with N7-methylguanosine sites and 741 RNA sequences without N7-methylguanosine sites | RNA N7-methylguanosine sites prediction |
| | M6A-BERT-Stacking | [51] | March 2023 | 12 transformer layers with 12 attention heads in each layer. | BERT-based | RNA sequence | RNA sequence with m6A sites and RNA sequence without m6A sites identified from MeRIP, m6A-seq, PA-m6A-seq, and miCLIP [52] | 11 datasets including 3000~ 16000 RNA sequences for each dataset | RNA m6A sites prediction |
| | Bert2Ome | [53] | May 2023 | 16 heads, 12 layers, and 1024 hidden units | BERT-based | RNA sequence | 2-O-methylation modification sites from RMBase database [54] | 215 positive, 215 negative instances for the training part and 46 positive, 114 negative instances for the testing part. | RNA 2-O-methylation prediction |
| | Rm-LR | [55] | Sep 2023 | 6 transformer | BERT- | RNA | Transcriptomic-wide | 20 different epi-transcriptome | Multiple types of RNA |
| | | | | encoder layers, with a hidden layer size of 512 and 16 attention heads | based | sequence | profiling data derived from the MultiRM, GEO, RMBase, RADAR [54, 56-58] | profiles based on various base resolution techniques | modifications prediction |
| | BertNDA | [59] | Nov 2023 | 8 layers | BERT-based | RNA sequence | miRNA-disease associations, lncRNA-disease associations, and miRNA-lncRNA associations [60-63] | 1000 positive pairs and 1000 negative pairs | ncRNA-Disease Association Prediction |
| | LncCat | [64] | Feb 2023 | - | BERT-based | RNA sequence | lncRNAs and protein-coding transcripts of five species [10, 63, 65] | 22960 coding transcripts and 21081 lncRNAs of human. 20707 coding transcripts and 10707 lncRNAs of mouse. 15891 coding transcripts and 4382 lncRNAs of zebrafish. 4693 coding transcripts and 5377 lncRNAs of wheat 20584 coding transcripts and 3897 lncRNAs of chicken | Identify lncRNA |
| | LSCPP-BERT | [66] | Dec 2023 | 4 identical layers and each layer is divided into two sublayers | BERT-based | RNA sequence | lncRNAs sequences from multispecies [67] | 593251 plant lncRNAs sequences | lncRNA-sORFs coding potential prediction |
| | CodonBERT | [68] | Oct 2023 | 12 layers of bidirectional transformer encoders, Each transformer layer with 12 self-attention heads | BERT-based | RNA sequence | mRFP Expression dataset; [69] Fungal expression dataset; E. coli proteins dataset; mRNA stability dataset; Tc-Riboswitches dataset [70] | Experimental data for protein expression (2308 low expression proteins, 2067 medium expression proteins, and 1973 high expression proteins, respectively); | mRNA properties prediction |
| | | | | | | | SARS-CoV-2 Vaccine degradation dataset | - | Vaccine expression prediction |
| | RNA-TorsionBERT | [71] | Jun 2024 | 18 layer | BERT-based | RNA sequence | PDB structure and removed the structures from the nonredundant Training | 4,267 structures with sequences from 11 to 508 nucleotides | RNA 3D structure prediction |
| | UNI-RNA | [72] | bioRxiv posted Jul 2023 | - | BERT-based | RNA sequence | non-coding RNA sequences from RNAcentral, nucleic acid data from NCBI's database, and genomic data from repositories such as Genome Warehouse[73-75] | 37,149 RNA structures 7,600 held-out real human 5'UTRs 3 million distinct UTR sequences | RNA secondary structure prediction RNA distance map prediction mRNA 5′-UTR mean ribosome load prediction Alternative |
| | | | | | | | | | polyadenylation isoform prediction<br>RNA splice site prediction<br>ncRNA classification<br>RNA modification prediction |
| | UTR-LM | [76] | Apr 2024 | Six-layer transformer with 16 multi-head self-attention. | Transformer-based | RNA sequence | Unlabeled 5' UTR sequences from three sources: the Ensembl database | 214,349 unlabeled 5' UTR sequences | Mean ribosome loading prediction.<br>mRNA expression level and translation efficiency prediction<br>Internal ribosome entry site identification<br>Attention-based motif detection |
| | 3UTRBERT | [77] | Aug 2024 | 12 identical Transformer components， Each layer contained a multi-head self-attention module and a position-wise fully connected feed-forward layer | BERT-based | RNA sequence | 3'UTR of human mRNAtranscript<br>eCLIP datasets<br>RNA localization data | 108 573 unique mRNA transcripts<br>1000 samples RNA binding samples with1000 sequences<br>17 023 mRNAs | mRNA subcellular localization prediction |
| | RNAErnie | [78] | May 2024 | 12-layer transformer and a hidden state dimension of 768 | Transformer-based | RNA sequence | non-coding RNA sequences from RNAcentral | | RNA grouping<br>RNA sequence classification<br>RNA–RNA interaction prediction<br>RNA secondary structure prediction |

## Supplementary Table 2. Detailed information of large language models for proteomic tasks

| Application area | Models | Ref | Publication time | Parameters | Architecture | Datasets | | | Downstream tasks |
|---|---|---|---|---|---|---|---|---|---|
| | | | | | | **Data type** | **Source** | **Size** | |
| Protein Large Language Models | MSA Transformer | [1] | Jul 2021 | 100M parameters model with 12 layers, 768 embedding size, and 12 attention heads | Transformer-based | MSAs | CAMEO [2] | 131 domains (129 evaluated) | Unsupervised contact prediction, supervised contact prediction, secondary structure prediction |
| | | | | | | | CASP13-FM [3] | 31 free modeling domains (from 25 targets) | |
| | | | | | | | trRosetta training set [4] | 15,051 MSAs and structures (14,842 used) | |
| | | | | | | Protein sequences | CB513 [5] | 513 protein sequences | |
| | | | | | | | Netsurf dataset [6] | 12,185 crystal structures obtained from the PDB [7] | |
| | UniRep | [8] | Dec 2019 | 18.2M parameters (a 1,900-hidden unit mLSTM with amino-acid character embeddings) | LSTM-based | Protein sequences | UniRef50 [9] | 24M | Predicting stability of naturally occurring and de novo designed proteins, Prediction of functional effects of single mutations in diverse proteins |
| | | | | | | | Mini protein dataset from [10] | 1,432 out of 5,570 test set and 1,416 out of 5,571 validation set | |
| | | | | | | | DMS dataset [11] | ~65,420 variants across the 8 proteins | |
| | | | | | | | avGFP dataset [12] | 32,400 variants derived from 27 homologs of avGFP (used) | |
| | TAPE | [13] | Dec 2019 | 38M parameters | (ResNet, Transformer, LSTM)-based | Protein sequences | Netsurf dataset [6] | 12,185 protein sequences | Secondary structure (SS) prediction (structure prediction task), contact prediction (structure prediction task), remote homology detection (evolutionary understanding task), fluorescence landscape prediction (protein engineering task), stability landscape prediction (protein engineering task) |
| | | | | | | | ProteinNet dataset [14] | ~332,283,871 protein sequences | |
| | | | | | | | DeepSF dataset [15] | Training set includes 16,712 proteins spanning 1,195 folds [16]; test datasets include 2,533 protein domains across 550 folds from SCOP 2.06 [16], a subset of SCOP | |
| | | | | | | | | 1.75 and the CASP dataset [17, 18] | |
| | | | | | | | avGFP dataset [12] | ~51,715 protein sequences | |
| | | | | | | | Dataset from [10] | ~ 46,800 protein sequences | |
| | ESM-1b | [19] | Dec 2020 | A model with ~650M parameters (33 layers) | Transformer-based | Protein sequences | SCOPe [20] | 15,297 protein sequences | Remote homology detection, prediction of secondary structure, long-range residue–residue contacts, mutational effect prediction, etc. |
| | | | | | | | CB513 [5] | 513 protein sequences; | |
| | | | | | | | CASP13 [21] | 431 domains | |
| | | | | | | | Envision (DMS dataset) [11] and DeepSequence [22] | Over 700,000 variant effect measurements from over 100 large-scale experimental mutagenesis datasets | |
| | ProtTrans | [23] | Aug 2021 | From millions to billions parameters (224M-11B) | (Transformer-XL, XLNet, BERT, Albert, Electra, T5)-based | Protein sequences | CB513 [5] | 513 protein sequences | Per-residue secondary structure prediction，per-protein localization & membrane prediction |
| | | | | | | | TS115 [5] | 115 protein sequences | |
| | | | | | | | CASP12 [24] | ~102 protein sequences | |
| | | | | | | | NEW364 [23] | 364 protein sequences | |
| | | | | | | | DeepLoc [25] | ~19,817 protein sequences | |
| | | | | | | | SCOPe 2.07 [26] | 14,323 protein sequences (non-redundant at PIDE < 40%) | |
| | SPRoBERTa | [27] | Sep 2022 | 12 Transformer encoder layers, the embedding size is 768, the feed-forward hidden units are 3072 and the attention heads are 12 | BERT-based | Protein sequences | Netsurf dataset [6] | 12,185 protein sequences | Secondary structure prediction, contact prediction, remote homology prediction, protein function prediction or Gene Ontology (GO) term prediction |
| | | | | | | | CB513 [5] | 513 protein sequences | |
| | | | | | | | CASP12 [24] | ~102 protein sequences | |
| | | | | | | | DeepSF dataset [15] | Consistent with the size used TAPE [13] | |
| | | | | | | | DeepFRI [28] | ~284,832 protein sequences with GO annotations | |
| | PromptProtein | [29] | Sep 2022 | 650M parameters with 33 layers and 20 attention heads. The embedding size is 1280 | Transformer-based | Protein sequences | EC dataset from [28] | 19,199 protein sequences | Enzyme commission and Gene Ontology prediction, stability landscape prediction, fluorescence landscape prediction, thermostability landscape prediction, adeno-associated virus (AAV) landscape prediction, GB1 landscape prediction, antibody-antigen affinity prediction |
| | | | | | | | GO dataset from [28] | 36,641 protein sequences | |
| | | | | | | | Protein engineering dataset from TAPE and FLIP [13, 30] | 68,965 protein sequences used in stability prediction, 54,025 used in fluorescence landscape prediction, 28,131 used in thermostability landscape prediction, 82,583 in AAV prediction, 8,733 in GB1 prediction | |
| | ProGen | [31] | Mar 2020 | 1.2B parameters. Sequence length is 512. The model has dimension d = 1028, inner dimension f = 512, 36 layers, and 8 heads per layer. Dropout with probability 0.1 follows the residual connections in each layer | Transformer-based | Protein sequences, Conditioning tags | Uniparc [32], UniprotKB [33], SWISS-PROT [34], TrEMBL [35], Pfam [36], and NCBI taxonomic information [37] | 281M | Controllable protein generation and two case study: completing VEGFR2 kinase domain, zero-shot fitness selection for protein GB1 |
| | Tranception | [38] | Jun 2022 | 700M parameters | Transformer-based | Protein sequences | Protein Gym (a DMS dataset) [38] | ~1.8M protein sequences | Fitness prediction |
| | ProtGPT2 | [39] | Jul 2022 | 738M parameters. The model consists of 36 layers with a model dimensionality of 1280. The architecture matches that of the previously released GPT2-large | GPT-based | Protein sequences | UniRef50 [40] | 10,000 protein sequences | Sequence dataset generation, homology detection, disorder prediction |
| | | | | | | | ProtGPT2 dataset [39] | Generated 10,000 protein sequences | |
| | ProteinBERT | [41] | Feb 2022 | 16M parameters. The model architecture consists of two almost parallel paths: one for local representations with | BERT-based | Protein sequences, Gene Ontology (GO) annotations | Secondary structure dataset from | 8,678 sequences (train) [13, 42] | (Secondary structure, disorder, Remote homology, fold classes, signal peptide, major PTMs, neuropeptide cleavage, |
| | | | | | | | Disorder dataset | 8,678 sequences (train) | |
| | | | | d=128 and the other for global representations with d=512 | | | from | [42] | fluorescence, stability) prediction |
| | | | | | | | Remote homology dataset from | 12,312 sequences (train) [13, 43, 44] | |
| | | | | | | | Fold classes dataset from | 15,680 sequences (train) [43, 44] | |
| | | | | | | | Signal peptide dataset from | 16,606 sequences (train) [45] | |
| | | | | | | | Major PTMs dataset from | 43,356 sequences (train) [46] | |
| | | | | | | | Neuropeptide cleavage dataset from | 2,727 sequences (train) [47, 48] | |
| | | | | | | | Fluorescence dataset from | 21,446 sequences (train) [12, 13] | |
| | | | | | | | Stability dataset from | 53,679 sequences (train) [10] | |
| | ProtST | [49] | Jan 2023 | Depends on the parameters of the chosen language model | Multi-models-based (Protein Language Models and Biomedical Language Models) | Protein sequences, property descriptions | ProtDescribe [50, 51] | 553,052 aligned pairs of protein sequence and property description | Protein localization prediction, fitness landscape prediction, protein function annotation (totally 11 downstream tasks) |
| | KeAP | [52] | Jan 2023 | Depends on the parameters of the chosen language model | Multi-cascade Bert-like network | Triplet in the format of (Protein, Relation, Attribute) | ProteinKG25 [53] | 5M with nearly 600k protein, 50k attribute terms, and 31 relation terms included | Amino acid contact prediction, protein homology detection, protein stability prediction, protein-protein interaction identification, protein-protein binding affinity prediction, and semantic similarity inference |
| | CaLM | [54] | Jan 2024 | 86M parameters. 12 transformer layers contain 12 attention heads, with dimension 768. Similar to architectures of ESM family | Transformer-based | Protein-coding DNA (cDNA) | cDNA dataset obtained from the European Nucleotide Archive with a timestamp of April 2022 | 9,858,385 cDNA sequences of seven model organisms | Melting point prediction, solubility prediction, subcellular localization prediction and function prediction |
| | | | | | | | Melting temperature dataset [30, 55] | - | |
| | | | | | | | Subcellular localization dataset [30, 56] | - | |
| | | | | | | | Solubility dataset [57] | - | |
| | | | | | | | Gene ontology dataset [58] | - | |
| | | | | | | | Transcriptomics dataset [59] | - | |
| | | | | | | | Proteomics dataset [60] | - | |
| | PLMSearch | [61] | Mar 2024 | ESM-1b (650M parameters) [19] and ProtT5-XL-UniRef50 (3B parameters) [23] | Transformer-based | Protein sequences | SCOPe40 [20, 62], New protein search test, Swiss-Prot [34], CATHS40 [63] | ~489,764 sequences for training | Homologous protein search |
| | DHR | [64] | Jul 2024 | 2 ESM-1b (650M parameters per encoder) [19] | Transformer-based | Protein sequences | UniRef [9], SCOPe [20, 62] | ~2 M | Protein homolog detection |
| Antibody Large Language Models | MHCRoBERTa | [65] | Dec 2021 | Model with 12multi-heads and 5 self-attention layers. | ROBERTa-based | Protein sequences | UniProtKB [66] | 565,254 protein sequences | Predicting the binding of peptide and major histocompatibility complex (MHC) |
| | | | | | | | Immune Epitope Database (IEDB) [67] | MHCclass I transmembrane proteins containing HLA-A(1,777 sequences)，HLA-B (2,100 sequences) and HLA-C(1,931 sequences) | |
| | BERTMHC | [68] | Jun 2021 | The model has 12 layers with 12 self-attention heads in each layer | BERT-based | Protein sequences | The data from Kamilla Kjærgaard Jensen [69] | 2,413additional MHC–peptide pairs covering 47 MHC class II alleles. | Predicting precisely the binding and presentation of peptides to major histocompatibility complex (MHC) alleles |
| | | | | | | | Immune Epitope Database (IEDB) [67] | 95,638 peptides | |
| | TCR-BERT | [70] | BioRxiv posted Nov 2021 | 12 stacked transformer blocks with 8 attention heads, utilizing a hidden representation dimensionality of 768 and featuring 12 transformer layers. | BERT-based | Protein sequences | Pan immune repertoire database (PIRD) [71] | 47,040 TRB sequences and 4,607 TRA sequences. | Antigen specificity classification |
| | | | | | | | VDJdb [72] | 58,795 human TCRs and 3,353 mouse TCRs. | |
| | | | | | | | TCRdb [73] | 139,00,913 TRB sequences of unknown antigen binding affinity. | |
| | Antiformer | [74] | July 2024 | he transformer encoder involves 12 stacking layers of transformer with the multi-head self-attention and feed forward network. | BERT-based | Protein sequences Gene expression | The OAS database[75] | 55 BCR-seq datasets containing 600 million sequences | Binding specificity prediction |
| | SC-AIR-BERT | [76] | May 2023 | Six standard transformer layers and each layer has four attention heads, the hidden representation dimensionality is 512 and the inter mediate representation dimensionality is 2048 | BERT-based | Protein sequences | VDJdb [72] | 23,358 unique paired TCRs | Binding specificity prediction |
| | | | | | | | Immune Epitope Database (IEDB) [67] | 18,662 paired TCR sand 589 paired BCRs | |
| | | | | | | | huARdb [77] | 612,077 high-confidence paired full-length α/β chains of | |
| | | | | | | | | TCR sequences | |
| | | | | | | | CoV-AbDab [78] | 1,105,906 paired antibody heavy/light chains | |
| | AbLang | [79] | Jun 2022 | Consists of three modules (Each of AbRep's 12 transformer blocks has 12 attenuated heads, an inner hidden size of 3072 and a hidden size of 768. From AbRep, the rescodings (768 values for each residue) are obtained. AbHead follows the design of RoBERTa's[80] head model, with a hidden size of 768) | BERT-based | Human antibody sequences | Observed Antibody Space (OAS) database [81] | Training sets of 14,126 724 heavy and 187,068 light sequences, and two evaluation sets of 100,000 heavy and 50,000 light sequences | Sequence specific predictions, residue specific predictions, amino acid predictions |
| | | | | | | | Data from [82] | 10,000 naive and 10,000 memory B-cell sequences | |
| | AntiBERTa | | Jul 2022 | 86M parameters. A 12-layer transformer model. Attention heads is 12, embedding dimension is 768, feedforward layer dimension is 3072 | BERT-based | Human antibody sequences | SAbDab [83] | Training/validation/test split of 720/90/90 | Trace the B cell origin of the antibody, quantify immunogenicity, predict the antibody's binding site |
| | | | | | | | BCR repertoire dataset [82] | - | |
| | | | | | | | TheraSAbDab [84] | 191 non-redundant therapeutic antibodies | |
| | EATLM | [85] | Jan 2023 | 86M parameters (12 layers, 12 heads, and 768 hidden states) | Transformer-based | Human antibody sequences | Dataset from [86] | Training/validation/test split of 15,128/3,242/3,242 | Accurate antigen-binding prediction， paratope prediction， B cell analysis， antibody discovery |
| | | | | | | | Paratope data from [87] | 1,662 CDR segments on 277 antibodies | |
| | | | | | | | Data from [88] | 88,094 sequences with 6 maturation stages | |
| | | | | | | | A subset of the OAS database [81] | Antibody sequences from 133 SARS-CoV-2 patients and 87 health persons | |

## Supplementary Table 3. Detailed information of large language models for drug-discovery tasks

| Application area | Models | Ref | Publication time | Parameters | Architecture | Datasets | | | Downstream tasks |
|---|---|---|---|---|---|---|---|---|---|
| | | | | | | Data type | Source | Size | |
| Predictions of Molecular Properties | SMILES-BERT | [1] | Sep 2019 | 6 Transformer encoder layers, the feed-forward hidden units are 1024 and the attention heads are 4 | BERT-based | SMILES | ZINC [2] | 18.69 M used, totally more than 35 M | LogP prediction, PM2 prediction, and molecular property prediction |
| | ChemBERTa | [3] | Oct 2020 | Implementation of RoBERTa uses 12 attention heads and 6 layers, resulting in 72 distinct attention mechanisms. | BERT-based | SMILES | PubChem [4] | Curated a dataset of 77 M unique SMILES from PubChem, divided this dataset into subsets of 100 K, 250 K, 1 M, and 10 M. | Binary classification prediction of barrier permeability properties, binary classification of clinical trial toxicity, whether the compound inhibits HIV replication for binary classification between active and inactive |
| | ChemBERTa-2 | [5] | Sep 2022 | Implementation of RoBERTa uses 12 attention heads and 6 layers, resulting in 72 distinct attention mechanisms. | BERT-based | SMILES | PubChem | Over a large corpus of 77 million SMILES strings | brain penetrability, toxicity, solubility, and on-target inhibition |
| | K-BERT | [6] | Apr 2022 | The hidden size of the transformer encoder is 768, and the number of the attention heads is 12. Six transformer encoders were used in KBERT. | BERT-based | SMILES | ChEMBL [7] | 1.8 M used | Related tasks on 15 drug-discovery-related datasets including carcinogenicity, respiratory toxicity and drug induced liver injury. |
| | MOLE-BERT | [8] | Apr 2023 | A 5-layer Graph Isomorphism Networks (GINs) whose hidden dimension is 300 | BERT-style, graph-based | Molecular graphs | ZINC15 [9] | 2 million molecules sampled from the dataset | Related tasks on 8 drug-discovery-related datasets including barrier permeability properties and clinical trial toxicity |
| Generation of Molecules | MolGPT | [10] | Oct 2021 | 6 M parameters. Each self-attention layer returns a vector of size 256 that is taken as input by the fully connected network. The hidden layer of the neural network outputs a vector of size 1024 and passes it through GELU activation layer. The final layer of the fully connected neural network returns a vector of size 256, that is then used as input for the next decoder block. MolGPT consists of eight such decoder blocks | GPT-based | SMILES | MOSES [11] and GuacaMol [12] | 1.9 M, 1.6 M | Generating molecules |
| Drug-Target Interaction | DTI-BERT | [13] | Jun 2022 | The proteins can be represented via 1024-D vectors (dimensionality of the features extracted by the ProtBert model). Drug molecular fingerprints are represented by 128-D vectors through semi decomposition process discrete wavelet transform (DWT). Secondly, the 1152-D vectors (a concatenation of protein sequence feature and drug feature) are fed into the feature extraction model to generate interaction information | BERT-based | Molecular fingerprints and protein sequence pairs | DrugBanks [14], BRENDA, SuperTarget, and KEGG BRITE [15] | 4,803 drug-target pairs in positive subsets 9,606 synthesized negative pairs | A pair belongs to an interactive drug-target pair or non-interactive drug-target pair |
| | TransDTI | [16] | Jan 2022 | Consist of SMILES-BERT and fine-tuned large protein models. | BERT-based | Molecular SMILES and protein sequence pairs | KIBA [17] , gold-standard external data sets from DTI-MLCD [18] | 30, 474 compounds, 961 targets and 61, 624 interactions | A three-classification based on binding affinity |
| | C2P2 | [19] | Jul 2022 | Consist of ChemBERTa and ESM model | BERT-based | Molecular SMILES and protein sequence pairs | STRING [20], STITCH [21], Davis [22], and PDBBind v2019 [23, 24] | Over 67.6 million proteins with over 20 billion protein–protein pairs, over 0.5 million chemicals with over 1.6 billion interactions, 30,056 interactions, 14,011 interactions | Binding affinity prediction |
| | Hyeunseok Kang et al. | [25] | Aug 2022 | Consist of ChemBERTa and ProtBERT model | BERT-based | Molecular SMILES and protein sequence pairs | BIOSNAP [26], DAVIS [22] and BindingDB [27] | 27,482 interactions, 11,103 interactions, 32,601 interactions | Binding affinity prediction |

## Supplementary Table 4. Detailed information of large language models for single-cell tasks

| Application area | Models | Ref | Publication time | Parameters | Architecture | Fine-tuning datasets | | | Downstream tasks |
|---|---|---|---|---|---|---|---|---|---|
| | | | | | | Data type | Source | Size | |
| Single-cell large language models | scBERT | [1] | Sep 2022 | six Performer encoder layers and ten heads for each layer, 200 dimensions of gene embedding using gene2vec | BERT-based | scRNA-seq | The Panglao dataset [2] | 209 human single-cell datasets comprising 74 tissues with 1,126,580 cells | Cell type annotation, novel cell type discovery, robustness to batch effects and model interpretability |
| | | | | | | | Zheng68k dataset [3] | 68,450 cells | |
| | | | | | | | Pancreas datasets [4-7] | - | |
| | | | | | | | MacParland dataset [8] | 8,444 cells | |
| | | | | | | | Heart datasets [9, 10] | 451,513 cells for pretraining and the 287,269 cells for benchmarking | |
| | | | | | | | Lung dataset [11] | 39,778 cells | |
| | | | | | | | Human Cell Atlas dataset [12] | 84,363 cells from 27 cell types among 15 major organs | |
| | scGPT | [13] | Feb 2024 | 12 stacked transformer blocks with 8 attention heads, 512 embedding sizes of the pre-trianed foundation model, 512 hidden sizes of the fully connected layer | GPT-based | scRNA-seq | CELLxGENE scRNA-seq human PBMC Collection [14] | 33 million human PBMC scRNA-seq samples | Gene network inference |
| | | | | | | | PBMC 10K dataset [15] | Two scRNA-seq data of 7,982 cells and 4,008 cells. | Multi-batch integration |
| | | | | | | | Immune Human dataset [16] | 33,506 cells | |
| | | | | | | | hPancreas dataset [4-7, 17, 18] | 10,600 cells in the reference set and 4,218 cells in the query set | Cell type annotation |
| | | | | | | | Adamson dataset [19] | 87 unique one-gene perturbations, each replicated in around 100 cells | Genetic perturbation prediction |
| | | | | | | | Norman dataset [20] | 131 two-gene perturbations and 105 one-gene perturbations. Each perturbation is replicated in around 300-700 cells. | |
| | | | | | | Single-cell multi-omics | 10X Multiome PBMC [21] | 9,631 cells | Multi-omic integration |
| | | | | | | | ASAP PBMC [22] | Four datasets each contain 5,023, 3,666, 3,517, and 4,849 cells respectively | |
| | CIForm | [23] | July 2023 | 64 attention heads | BERT-based | scRNA-seq | Pancreas datasets [4-7] | - | Cell type annotation |
| | | | | | | | Immune datasets [24-26] | - | |
| | | | | | | | Brain datasets [27-29] | - | |
| | | | | | | | Tabula Muris dataset [30] | Nearly 100000 cells from 20 organs and tissues | |
| | | | | | | | Zheng68k dataset [3] | 68,450 cells | |
| | | | | | | | ZhangTdataset [31] | 8,530 cells from 20 subtypes | |
| | | | | | | | Allen mouse brain dataset [32] | 12832cells | |
| | TOSICA | [17] | Jan 2023 | - | BERT-based | scRNA-seq | human pancreas (hPancreas) [4-7] | 10,600 cells for training and 4,218 for testing | Cell type annotation, new cell type discovery and batch correction |
| | | | | | | | human bone (hBone, GSE152805) [33] | 14,615 cells for training and 11,525 for testing | |
| | | | | | | | human artery (hArtery, GSE159677) [34] | 10,960 cells for training and 35,399 for testing | |
| | | | | | | | mouse brain (mBrain) [28-30, 35] | 48,801 cells for training and 7,394 for testing | |
| | | | | | | | mouse pancreas (mPancreas, GSE132188) [36] | 25,465 cells for training and 10,886 for testing | |
| | | | | | | | mouse atlas (mAtlas, GSE132042) [37] | 78,672 cells for training and 277,541 for testing | |
| | scTransSort | [38] | Mar 2023 | 12 layers of transformer | BERT-based | scRNA-seq | human cell atlas dataset | 295,805 cells from 35 tissues | Cell type annotation |
| | | | | | | | mouse cell atlas dataset | 105,148 cells from 26 tissues and 103,148 cells from 26 tissues | |
| | | | | | | | data processed by Shao X et al. [39] | - | |
| | TransCluster | [40] | Oct 2022 | 5 attention heads | improved Transformer model | scRNA-seq | Human training data [41] | - | Cell type Identification |
| | | | | | | | The Shao dataset [39] | - | |
| | | | | | | | The Baron dataset [4] | - | |
| | Geneformer | [42] | May 2023 | six transformer encoder units, input size of 2,048, 256 embedding dimensions, four attention heads per layer and feed forward size of 512 | BERT-based | scRNA-seq | iPSC differentiation data | Assayed in parallel on the Drop-seq (single cell) or DroNc-seq (single nucleus) platform | Batch effect removal, cell type annotation |
| | | | | | | | Huggingface Dataset [43] | a largescale pretraining corpus, Genecorpus-30M, comprising 29.9 million human single-cell transcriptomes | discover key network regulators and candidate therapeutic targets |
| | tGPT | [44] | May 2023 | 8 transformer decoder blocks with 1024 hidden units and 16 attention heads | GPT-based | scRNA-seq | Human Cell Atlas Census of Immune Cells (HCA) [45] | 282,588 Bone marrow cells from 64 healthy donors in Human Cell Atlas (HCA) project | single-cell clustering, inference of developmental lineage |
| | | | | | | | Human cell Landscape (HCL) [41] | 586,135 human cells | |
| | | | | | | | Tabula Mursi dataset [30] | 54,865 cells | |
| | | | | | | | Macaque Retina dataset [46] | 124,965 cells | |
| | | | | | | Bulk RNA-seq | The Cancer Genome Atlas (TCGA) [47] | 9,318 bulk samples | interrogation of feature representation of bulk tissues in relation to genomic alterations, prognosis and treatment response of immunotherapy |
| | | | | | | | Genotype-Tissue Expression Project (GTEx) [48] | 11,688 bulk samples | |
| | DeepMAPS | [49] | Feb 2023 | - | BERT-based, graph transformer | scRNA-seq | Multiple scRNA-seq data [4, 16] | Three scRNA-seq datasets with 20,125 cells, 14,878 cells and 16,382 cells | Cell clustering, infer cell-type-specific biological networks from scMulti-omics data |
| | | | | | | Single-cell multi-omics | CITE-seq data [50] | Three CITE-seq datasets with 25,171 cells, 32,029 cells and 16,750 cells | |
| | | | | | | | scRNA-ATAC-seq data | Four scRNA-ATAC-seq datasets with 3,009 cells, 11,898 cells, 3,233 cells and 10,970 cells | |
| | scMVP | [51] | Jan 2022 | 8 self-attention heads and each head takes 16-dimension feature | Transformer-based | Paired scRNA-seq and scATAC-seq data | sci-CAR cell line dataset [52] | 293T cell line, 3T3 cell line, 293T/3T3 cell mixture, and A549 cell line treated with dexamethasone (DEX) for 0 h, 1 h, and 3 h | dimensionality reduction, cell clustering, and developmental trajectory inference and generate separate imputations for differential analysis and cis-regulatory element identification |
| | | | | | | | Paired-seq cell line dataset [53] | derived from HEK293, HepG2, and their cell line mixture | |
| | | | | | | | SNARE-seq cell line dataset [54] | 5,081 cells | |
| | | | | | | | SHARE-seq [55] | 67,418 cells | |
| | scTranslator | [56] | BioRxiv posted July 2023 | 8-headed attention mechanism in each sub-layer, 117 million parameters. | Transformer-based | Bulk datasets | The Cancer Genome Atlas (TCGA) data [57-60] | 31 cancer types and 18,227 samples in total | translate single-cell transcriptome to proteome, predict protein abundance |
| | | | | | | | The data from Clinical Proteomic Tumor Analysis Consortium (CPTAC) [61-67] | | |
| | | | | | | | The dataset from Broad Institute [68, 69] | | |
| | | | | | | | The dataset from Memorial Sloan Kettering Cancer Center (MSKCC) [70] | | |
| | | | | | | Single-cell | The Seurat v4 PBMCs dataset | 161,764 human peripheral blood | |
| | | | | | | datasets | [71] | mononuclear cells | |
| | | | | | | | The REAP-seq PBMCs dataset [72] | 4,330 PBMCs with simultaneous measurements of 44 proteins and 21,005 transcriptome genes | |
| | | | | | | | The CITE-seq CBMCs dataset [73] | simultaneous measurements of 13 cellular proteins and 16,508 transcriptome genes. This dataset includes 8,005 cord blood mononuclear cells (CBMCs) | |
| | | | | | | | The single-cell pan-cancer dataset [74] | 65,698 myeloid cells with only single-cell transcriptome data, involving 15,844 genes | |
| | scFoundation | [75] | June 2024 | 100 million parameters | BERT-based | Single-cell datasets | Baron dataset [4] | - | gene expression enhancement, tissue drug response prediction, cell clustering, single-cell drug response classification, and single-cell perturbation prediction |
| | | | | | | | Zheng68K dataset [3] | 68,450 cells | |
| | | | | | | | Cancer drug response dataset [76] | - | |
| | | | | | | | Single cell drug response classification dataset | - | |
| | | | | | | | Perturbation dataset [77] | - | |
| | scMoFormer | [78] | Oct 2023 | Graph transformer | Transformer-based | Single-cell multi-omics | Joint measurements of gene expression and surface protein levels datasets from the NeurIPS multimodal single-cell integration competition of the year 2021 [25] and 2022 | - | use gene expression (RNA) to predict surface protein level, protein levels to gene expression, gene expression to chromatin accessibility and chromatin accessibility to gene expression |
| | GeneCompass | [79] | Sep 2024 | 12-layer transformer framework, 100 million parameters | BERT-based | single-cell transcriptomes | CHIP-Atlas related to PBMC cells on GSE43036 [80] | - | Gene embedding analysis, Gene expression profiling prediction |
| | | | | | | | human multiple sclerosis | - | cell type annotation |
| | | | | | | | (hMS), lung (hLung) and liver (hLiver) datasets, and mouse brain (mBrain), lung (mLung) and pancreas (mPancreas) datasets. | | |
| | | | | | | | Immune Human [16] | 33,506 cells | GRN inference |
| | | | | | | | dataset provided by Srivatsan et al. [81] | - | Drug dose response prediction |
| | | | | | | | predefined dosage-sensitive and nonsensitive gene datasets [42] | - | Gene dosage sensitivity predictions. |
| | | | | | | | Norman dataset [20] | 131 two-gene perturbations and 105 one-gene perturbations. Each perturbation is replicated in around 300-700 cells. | In silico perturbation |
| | | | | | | | mouse embryonic stem cells (ESCs) [82] | - | In silico quantitative perturbation |
| | scMulan | [83] | BioRxiv posted Jan 2024 | 24-layer transformer, 368 million parameters | GPT-based | single-cell transcriptomic data | AHCA_BoneMarrow [12] | 3,000 bone marrow cells from a single adult donor | zero-shot cell type annotation |
| | | | | | | | Simonson2023 [84] | 60,345 cells from 8 human left ventricle samples | |
| | | | | | | | Suo2022 [85] | 140,000 liver cells from 14 fetal donors at various developmental stages | |
| | | | | | | | Intestine_HCL_55k dataset [41] | 55,214 intestinal cells across 24 cell types | |
| | | | | | | | Immune cell dataset [86] | 274,346 cells spanning 18 batches | batch integration |
| | | | | | | | Lung dataset [16] | 32,472 cells from 16 donors | |
| | | | | | | | six organs within the hECA-10M dataset [83] | 3,000 conditions | conditional cell generation |
| | UCE | [87] | BioRxiv posted Nov 2023 | a 33-layer model consisting of over 650 million parameters | BERT-based | single-cell gene expression dataset | Tabula Sapiens v2 dataset | human data from 581,430 cells, 27 tissues, batches and 162 unique cell types | zero-shot embedding of new datasets |
| | | | | | | | a dataset of green monkey lymph node and lung cells [88] | 17 cell types | Cell type embedding for new species |
| | | | | | | | naked mole rat spleen and circulating immune cells [89] | 24 cell types | |
| | | | | | | | two distinct chicken datasets (chick retina [90] and | 15 cell types in chicken heart dataset | |
| | | | | | | | developing chick heart [91]) | | |
| | | | | | | | mouse renal cells [92] | | decode the function of newly discovered cell types |
| | CellLM | [93] | arXiv posted June 2023 | Performer model, 10 layers, 16 attention heads, over 50 million parameters | BERT-based | scRNA-seq data | Zheng68k dataset [3] | 68,450 human peripheral blood mononuclear cells (PBMCs) with 11 highly related cell types | Cell type annotation |
| | | | | | | | The pancreas Baron dataset [4] | 8,562 cells categorized into 13 different cell types | |
| | | | | | | | Human lung cancer cells (GSE149383) [94] | 2,739 cells | Single-cell drug sensitivity prediction |
| | | | | | | | human oral squamous cancer cells (GSE117872) [95-97] | 1,302 cells | |
| | | | | | | Single-omics cell line data | cell lines integrating from CCLE [98] and GDSC [99] | 555 cell lines and 223 drugs | Single-omics cell line drug sensitivity prediction |
| | scCLIP | [100] | Oct 2023 | Vanilla transformers | transformer-based encoders | Paired scRNA and scATAC-seq data | Fetal Atlas [101, 102] | 377, 134 cells | integration of multi-modal single-cell sequencing data |
| | | | | | | | Brain [103] | - | |
| | iSEEEK | [104] | Jan 2022 | 8 transformer layers each with 576 hidden units and 8 attention heads | BERT-based | single-cell expression data | PBMCs [105] | 43,073 cells | Cell clustering |
| | | | | | | | Human Cell Atlas Census of Immune Cells (HCA) [45] | 282,588 Bone marrow cells from 64 healthy donors in Human Cell Atlas (HCA) project | |
| | | | | | | | Tabula Muris dataset [30] | Nearly 100000 cells from 20 organs and tissues | |
| | | | | | | | Zheng68k dataset [3] | 68,579 cells | |
| | | | | | | | the dataset of FACS-sorted CD4/8+ T cells [31, 106, 107] | 12,670 CD4+ and 9,012 CD8+ T cells | Identify gene–gene interaction networks |
| | CellPLM | [108] | BioRxiv posted Oct 2023 | over 80 million parameters | BERT-based | scRNA-seq and spatially-resolved transcriptomic (SRT) data | dataset from Li et al. [109] | 48, 082 cells | zero-shot clustering |
| | | | | | | | PBMC 5K and Jurkat from 10x Genomics | 33,538 cells in PBMC 5K and 32,738 cells in Jurkat | scRNA-seq denoising |
| | | | | | | | two spatial transcriptomic datasets at single-cell resolution, i.e., Lung2 and Liver2 | 836,739 cells in Lung2 and 598,141 cells in Liver2 | spatial transcriptomic imputation |
| | | | | | | | hPancreas [17] and Multiple Sclerosis (MS) [110] | - | cell type annotation |
| | | | | | | | the Adamson Perturb-Seq | 87 one-gene perturbations in the | perturbation |
| | | | | | | | dataset [19] and the Norman Perturb-Seq dataset [20] | Adamson Perturb-Seq dataset, and 131 two-gene perturbations and 105 one-gene perturbations in the Norman Perturb-Seq dataset | prediction |
| | scGREAT | [111] | Feb 2024 | - | Transformer-based | single-cell transcriptomics | human embryonic stem cells (hESC) (GSE75748) [112] | - | gene regulatory network inference |
| | | | | | | | human mature hepatocytes (hHEP) (GSE81252) [113, 114] | - | |
| | | | | | | | mouse dendritic cells (mDC) (GSE48968) [115] | - | |
| | | | | | | | mouse embryonic stem cells (mESC) (GSE98664) [116] | - | |
| | | | | | | | mouse hematopoietic stem cells with erythroid-lineage (mHSC-E) mouse hematopoietic stem cells with granulocyte-monocyte-lineage (mHSC-GM) mouse hematopoietic stem cells with lymphoid-lineage (mHSC-L) (GSE81682) [117] | - | |
| | BioFormers | [118] | bioRxiv posted Dec 2023 | An 8-layer transformer encoder model with 8 self-attention heads per layer and a hidden state dimension of 512 | BERT-based | scRNA-seq | PBMC [15] | 7,982 cells and 3,346 genes | Cell clustering and identification |
| | | | | | | | PBMC 4k and 8k [119] | 11,990 cells and 2,000 HVGs | Gene expression prediction |
| | | | | | | | Perturb-seq dataset [19] | 87 single-gene perturbations, with ~100 cells per perturbation and a control set of at least 7,000 unperturbed cells | genetic perturbation prediction and gene network inference |
| | scPRINT | [120] | bioRxiv posted July 2024 | 2M to 100M parameters, 4-layer transformer encoder model with 2 self-attention heads per layer | BERT-based | scRNA-seq | three test datasets of kidney, retina, and colon tissues [121-123] | comprising 26 cell types | gene network inference, denoising, batch effect correction, and cell label prediction |
| | | | | | | | perturb-seq [124] and ChIP-seq [125] | - | |
| | | | | | | | 3 test datasets of ciliary body, colon, and retina [122, 126, 127] | - | |
| | | | | | | | premalignant neoplasms from | - | |
| | | | | | | | human prostate tissues [128] | | |
| | ScRAT | [129] | Feb 2024 | one-layer transformer encoder model with 8 self-attention heads | Transformer-based | scRNA-seq | COMBAT [130] and Haniffa datasets [131] | 835,937 and 528,438 cells | disease diagnosis |
| | | | | | | | SC4 with COVID samples [132] | 501,943 cells for severity prediction<br>1,289,496 cells for stage prediction | predict severity and stage |
| | CancerFoundation | [133] | bioRxiv posted Nov 2024 | Six transformer layers, 10.8 million parameters | BERT-based | scRNA-seq | glioblastoma dataset [134] | four distinct malignant cell states | batch integration |
| | | | | | | | CCLE [98] and GDSC [99] | - | Drug response prediction |
| | | | | | | bulk RNA-seq data | TCGA data [135, 136] | 21 cancer types | Survival prediction |
| | mcBERT | [137] | bioRxiv posted Nov 2024 | 12 blocks, each with 12 attention heads | BERT-based | scRNA-seq | Heart [9, 84, 138-141] | Refer to Table 1 in [137] | patient-level representation, disease clustering, phenotypical interpretation/batch effect removal |
| | | | | | | | Kidney [142-146] | | |
| | | | | | | | PBMC [130, 147, 148] | | |
| | | | | | | | Lung [149] | | |
| | Nicheformer | [150] | bioRxiv posted Oct 2024 | The transformer block leverages 12 transformer encoder units19,27 with 16 attention heads per layer and a feed-forward network size of 1,024 to generate a 512-dimensional embedding of the pretraining dataset, resulting in altogether 49.3M parameters. | BERT-based | single-cell and spatial transcriptomics data | MERFISH mouse brain [151] | 4.3 million cells across 59 tissue sections | Label prediction (cell type, niche and region labels) |
| | | | | | | | CosMx human liver [152] | 332,877 healthy cells and 460,441 cancer cells | niche label prediction (healthy data only), niche composition prediction |
| | | | | | | | CosMx human lung [152] | five different donors (301,611, 89,975, 227,110, 71,304 and 81,236 cells, respectively) | niche composition prediction |
| | | | | | | | Xenium human lung from 10X Genomics | 295,883 healthy cells and 531,165 cancer cells | neighborhood density prediction |
| | | | | | | | Xenium human colon from 10X Genomics | 275,822 healthy cells and 587,115 cancer cells | |
| | SpaFormer | [153] | arXiv posted Feb 2023 | 2 layers and 8 heads | Transformer-based | Spatial transcriptomics data | Lung 5 data generated by the CosMX platform [152] | 99,656 cells | spatial transcriptomic imputation |
| | | | | | | | Kidney 1139 data generated by the CosMX platform [152] | 61,283 cells | |
| | | | | | | | Liver normal generated by the CosMX platform [152] | 305,730 cells | |